\documentclass[aps,prb,twocolumn,preprintnumbers,amsmath,amssymb,colorlinks,showpacs,floatfix]{revtex4}

\usepackage{graphicx}
\usepackage{bm}
\usepackage{color}
\usepackage{hyperref}

\newcommand {\bcv} {BaCo$_2$V$_2$O$_8$}
\newcommand {\bacovo} {BaCo$_2$V$_2$O$_8$ }

\begin{document}

\title{Field-induced magnetic behavior in quasi-one-dimensional Ising-like antiferromagnet \bcv: A single-crystal neutron diffraction study}

\author{E. Can\'evet}
\affiliation{Institut Laue Langevin, BP 156, 38042 Grenoble cedex 9, France}
\affiliation{Universit\'e Joseph Fourier--Grenoble I, BP 53, 38041 Grenoble cedex 9, France}

\author{B. Grenier} 
\email[Corresponding author. Electronic address: ]{grenier@ill.fr}
\affiliation{SPSMS, UMR-E 9001, CEA--INAC/UJF--Grenoble I, laboratoire MDN, 17 rue des martyrs, 38054 Grenoble cedex 9, France}

\author{M. Klanj\v{s}ek}
\affiliation{Jozef Stefan Institute, Jamova 39, SI-1000 Ljubljana, Slovenia}
\affiliation{EN-FIST Centre of Excellence, Dunajska 156, SI-1000 Ljubljana, Slovenia}

\author{C. Berthier and M. Horvati\'c}
\affiliation {Grenoble High Magnetic Field Laboratory, CNRS, BP 166, 38042 Grenoble Cedex 9, France }

\author{V. Simonet and P. Lejay}
\affiliation{Institut N\'{e}el, CNRS/UJF--Grenoble I, BP 166, 38042 Grenoble Cedex 9, France}

\date{\today}

\begin{abstract}
\bacovo is a nice example of a quasi-one-dimensional quantum spin system that can be described in terms of Tomonaga-Luttinger liquid physics. This is explored in the present study where the magnetic field-temperature phase diagram is thoroughly established up to 12~T using single-crystal neutron diffraction. The transition from the N\'eel phase to the incommensurate longitudinal spin density wave (LSDW) phase through a first-order transition, as well as the critical exponents associated with the paramagnetic-N\'eel phase transition, and the magnetic order both in the N\'eel and in the LSDW phase are determined, thus providing a stringent test for the theory.\end{abstract}

\maketitle


\section{Introduction}
\label{sec:I}

One dimensional (1D) antiferromagnetic spin chains have attracted a lot of attention in the last decades, both theoretically and experimentally. A strong quantum character of these systems implies that they can show many fascinating phenomena, not existing in systems of higher dimension. Also, many magnetic properties can be exactly calculated, owing to their low dimensionality. More recently, magnetic field effects have been particularly focused on, since the application of a magnetic field can bring new behaviors, such as magnetization plateaus, Bose-Einstein condensation, or novel field-induced magnetic orderings.\cite{Giamarchi2008} Such novel phenomena were indeed observed respectively in {\it e.g.} the following systems of weakly-coupled spin chains: azurite containing distorted diamond spin-1/2 chains,\cite{Kikuchi2005, Aimo2009} Cu(NO$_3$)$_2\cdot$2.5H$_2$O containing alternating spin-1/2 chains,\cite{Grenier2007} and $Ni(C_5H_{14}N_2)_2N_3(PF_6)$, commonly abbreviated NDMAP, containing anisotropic Haldane spin-1 chains.\cite{Zheludev2004} A great deal of experimental work has been performed on spin-1/2 and spin-1 quantum 1D systems, but very little on systems of higher spin.

The quasi-1D spin-3/2 screw chain antiferromagnet \bacovo is particularly interesting since it possesses a strong anisotropy, some sizable frustration, and, in spite of a larger spin, strong quantum effects. Namely, the low-temperature magnetic state of the Co$^{2+}$ ion ($3d^7$) in a distorted octahedral environment, as it is the case in this compound, is described by a highly anisotropic effective spin $S=1/2$.\cite{Abragam1951} As a result, quantum fluctuations play an important role in this compound. 

In zero magnetic field, \bacovo exhibits upon cooling a transition from a 1D magnetic behavior to a 3D antiferromagnetic ordering at $T \sim 5$~K, as first evidenced by He {\it et al.}\cite{He2005} from magnetic and specific heat measurements. This system also presents a strong magnetic anisotropy, as shown from a twice larger susceptibility when the field is applied along the $c-$axis (chain direction), as compared to the one in a field applied perpendicular to the $c-$axis.\cite{He2005,He2006} All these results indicate that the chain $c-$axis is the magnetic easy axis in this compound. From magnetization measurements performed in a field applied along the $c-$axis, a field-induced order-to-disorder transition occurring above 1.8~K (critical field $H_c \simeq 3.8$~T at 1.8~K), instead of a spin-flop transition, was then evidenced,\cite{He2005b} before the saturation is reached at $H_s=22.7$~T.\cite{Kimura2006} Kimura {\it et al.}\cite{Kimura2006,Kimura2007} attributed this order-to-disorder transition to the quantum phase transition characteristic of the $S=1/2$ 1D XXZ antiferromagnet with Ising-like anisotropy model, whose Hamiltonian is given by: 
\begin{equation}
\mathcal{H} = J \sum_i \{ S_i^zS_{i+1}^z + \epsilon (S_i^xS_{i+1}^x+ S_i^yS_{i+1}^y)\} - g\mu_B \sum_i S_i^z\,H,
\end{equation}
where $S_i^\alpha$ (with $\alpha=x,\,y~{\rm or}~z$) denotes the $\alpha$-th component of the spin residing at the site $i$ of the chain, $J>0$ is the antiferromagnetic exchange interaction along the chain, $\mu_B$ is the Bohr magneton, $H$ is the longitudinal applied magnetic field ($H \parallel c-$axis), $g$ is the electron $g$-factor along the $c-$axis, and $\epsilon$ is an anisotropic parameter ($\epsilon < 1$ for the Ising-like case). The authors obtained the following parameters from the fit of the magnetization curves measured up to 55~T for $H\parallel c$: $J/k_B=65$~K ($k_B$ is the Boltzmann constant), $\epsilon \simeq 0.46$, and $g=6.2$.\cite{Kimura2006,Kimura2007}

An important concept of 1D quantum spin systems in their gapless phase is the Tomonaga-Luttinger liquid (TLL).\cite{Haldane1980} The transverse and longitudinal spin-spin correlation functions decay with the distance $r$ between the spins asymptotically as power laws, so that they depend on the TLL exponent $\eta$ as follows:
\begin{equation}
\left< S_0^xS_{r}^x \right> \simeq (-1)^r\,|r|^{-\eta}
\label{transverse} 
\end{equation}
and
\begin{equation}
\left< S_0^zS_{r}^z \right> - \left<S^z\right>^2 \simeq \cos(2k_Fr)\,|r|^{-1/\eta},
\label{longitudinal}
\end{equation}
where $\left<S^z\right>$ is the $z-$component of the spin due to the magnetic field $H$ applied parallel to the $c-$axis ($0 \leq \left<S^z\right> \leq 1/2$) and $k_F$ is the Fermi wave number defined by
\begin{equation}
k_F=\pi(1/2-\left<S^z\right>).
\label{FermiWaveVector}
\end{equation}
Mapping the spin chain onto a 1D system of spinless fermions using the Jordan-Wigner transformation,\cite{Giamarchi2004} this equation simply relates the filling of the fermion band $1/2-\left<S^z\right>$ with the wavevector $2k_F$ connecting both points of the Fermi surface. Alternatively, the transverse component of the correlation function can be said to decay with the exponent $\eta_x = \eta_y = \eta$, while the longitudinal component decays with the exponent $\eta_z = 1/\eta$. Relation~(\ref{FermiWaveVector}) implies that the value of $k_F$ can be tuned by the applied magnetic field $H$, through $\left<S^z\right>$. The longitudinal correlation function is thus incommensurate with the chain, while the transverse correlation function always appears at the wave number $\pi$, so that it is staggered.

When spin chains are weakly coupled, a long-range order of the spins sets in at low enough temperature. For the isotropic Heisenberg case ($\epsilon = 1$), the relation $\eta < 1$ ({\it i.e.}, $\eta_x < \eta_z$) is always satisfied in the presence of a magnetic field, and thus the transverse fluctuation of Eq.~(\ref{transverse}) is dominant, implying a long-range ordering of the $xy$ spin components (N\'eel order). However, for the Ising-like case, the relation $\eta > 1$ ($\eta_x > \eta_z$) holds above $H_c$ in the low field region of the gapless phase indicating a dominance of the longitudinal incommensurate correlation [see Eq.~(\ref{longitudinal})], before $\eta < 1$ ($\eta_x < \eta_z$) becomes satisfied at higher field: an inversion of the TLL exponents $\eta_x$ and $\eta_z$ occurs at some critical field $H^*$. Therefore, a long-range ordered phase with a longitudinal incommensurate structure is predicted theoretically between $H_c$ and $H^*$ before the establishment of a transverse staggered ordered phase at $H > H^*$. Based on Eq.~(\ref{longitudinal}), the magnetic structure propagation vector of the longitudinal incommensurate order is expected to be:
\begin{equation}
k=2k_F.
\label{PropWaveVector}
\end{equation}
As $\left<S^z\right>$ varies with the field, the propagation vector follows this variation. 

After the magnetization measurements of He {\it et al.}~\cite{He2005b} at 1.8~K, Kimura {\it et al.}\cite{Kimura2008} performed high-field specific heat measurements at 200~mK, which evidenced a new magnetic field-induced ordered state above $H_c \simeq 3.9$~T. The authors suggested that this field-induced magnetic phase exhibits the novel type of longitudinal incommensurate order described above, which is caused by the quantum effect inherent in the $S=1/2$ quasi-1D Ising-like antiferromagnet. This was then further explained by Okunishi and Suzuki,\cite{Okunishi2007} who calculated the TLL exponent $\eta$ by the Bethe ansatz integral equations, using the values of $J$, $\epsilon$ and $g$ determined from the magnetization curves and the interchain coupling $J'/k_B=0.09$~K estimated from the mean-field theory. They found that $\eta > 1$ between $H_c \simeq 3.9$~T and $H^*\sim15$~T in \bcv.\cite{Okunishi2007} This theoretical work strongly suggests that the field-induced transition observed below 1.8~K by Kimura {\it et al.}\cite{Kimura2008} corresponds to a N\'eel-longitudinal incommensurate order transition, or, more precisely, to a transition toward an incommensurate longitudinal spin density wave (LSDW). Okunishi and Suzuki also calculated the transition temperatures for the incommensurate and transverse staggered orders, as a function of the applied field, and obtained a very good agreement with the experimental observation for the low-field phase boundary of the longitudinal incommensurate ordering.\cite{Okunishi2007}

Soon after, this novel type of field-induced incommensurate order was further hinted by neutron diffraction.\cite{Kimura2008b} The magnetic 4\,0\,3 reflection was followed up to a magnetic field of 5~T, and was shown to split above $H_c\simeq 3.9$~T into incommensurate satellites at 4\,0\,3$\pm \xi$. The field dependence of $\xi$ above $H_c$ was compared to the theoretical one but only in a narrow field range of 1~T.

Very recently, Kawasaki {\it et al.}\cite{Kawasaki2010} determined the magnetic structure of \bacovo in zero magnetic field from a powder neutron diffraction experiment: the 2.18$\mu_B$ magnetic moments are aligned along the $c-$axis, with an antiferromagnetic arrangement along $c$ and along one of the crystallographic axes of the square basal plane ($a$ or $b$), and with a ferromagnetic arrangement along the other axis ($b$ or $a$).

\begin{figure*}
\begin{minipage}{.9\linewidth}
\includegraphics[width=14.5cm]{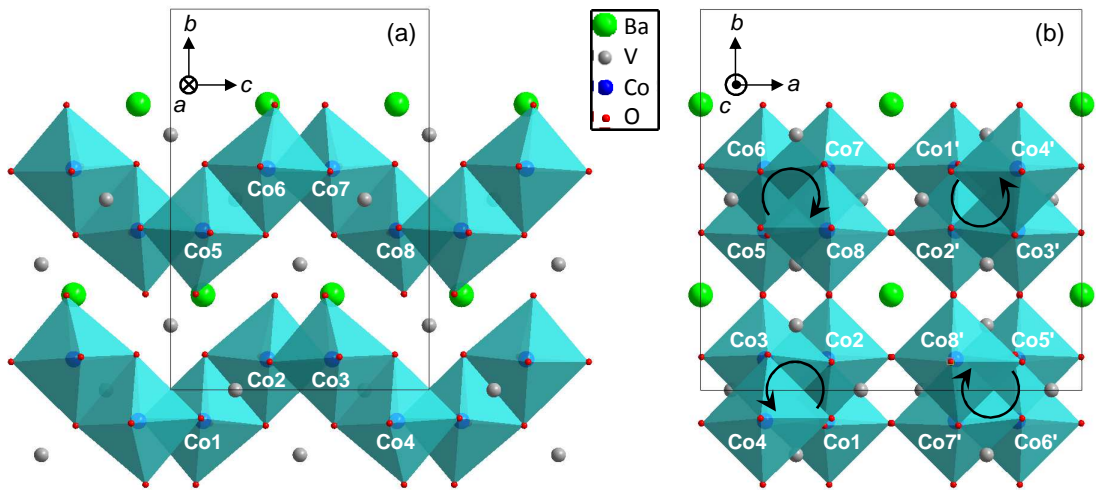}
\end{minipage}
\caption{(Color online) Crystal structure of \bcv. (a) Projection in the $(b, c)$ plane: for sake of clarity, only the two chains with $0 \leq x \leq 0.5$ and $-0.25 \leq y \leq 0.75$ are represented for $-0.62 < z < 1.62$. (b) Projection in the $(a, b)$ plane of the four chains of the unit cell ($0 \leq x \leq 1$, $-0.25 \leq y \leq 0.75$, $-0.15 < z < 1.15$): the arrows indicate the sense of rotation of the screw chains on increasing $z$.}
\label{StrucNuc}
\end{figure*}

In this paper, we present a complete exploration of the magnetic field-temperature $H$-$T$ phase diagram, up to 12~T and down to 50~mK, by means of single-crystal neutron diffraction, yielding various new and important results. First, besides the overall agreement of our phase diagram with that of Kimura {\it et al.}\cite{Kimura2008} (see Fig.~\ref{DiagHT}), the hysteretic behavior of the N\'eel-LSDW phase transition was carefully studied and a phase boundary marking the collapse of the LSDW phase, not predicted by the theoretical model of Okunishi and Suzuki,~\cite{Okunishi2007} was evidenced at around $9$~T. Second, a precise study of the critical exponents was performed in the low field part ($0 \leq H < H_c\simeq 3.9$~T) of the phase diagram. Third, the antiferromagnetic structure was determined in zero field, in good agreement with the one recently determined from powder neutron diffraction,\cite{Kawasaki2010} and the structure of the field-induced LSDW phase was determined for the first time. Last, a very detailed survey of the propagation vector and of the intensity of chosen magnetic reflections was performed up to 9~T, yielding an excellent agreement with the predicted field dependence of the incommensurate modulation over a large field range.

The paper is organized as follows: after the Introduction, the crystal structure and the experimental procedures are described in Sec.~\ref{sec:II}. In Sec.~\ref{sec:III}, the results are then presented and analyzed. In Sec.~\ref{subsec:IIIA}, a characterization of the sample by means of specific heat measurements is presented, followed by the presentation of the neutron scattering measurements. The determination of the nuclear and magnetic structures at $H=0$ are presented in Secs.~\ref{subsec:IIIB} and \ref{subsec:IIIC}, respectively. In Sec.~\ref{subsec:IIID}, the temperature and magnetic field scans are presented, including a study of the critical exponents and a careful survey of the incommensurate magnetic peak position as a function of the magnetic field, yielding the determination of the $H$-$T$ phase diagram. In Sec.~\ref{subsec:IIIE}, the magnetic structure in the LSDW phase is presented. Section~\ref{sec:IV} is devoted to a discussion, followed by a conclusion in Sec.~\ref{sec:V}.


\section{Experimental}
\label{sec:II}

\subsection{Crystal structure and growth}

\bacovo crystallizes in the centrosymmetric tetragonal body-centered $I4_1/acd$ (No. 142) space group, with $a=12.444$ \AA, $c=8.415$ \AA, and eight chemical formulas per unit cell.\cite{Wichmann1986} The 16 magnetic Co$^{2+}$ atoms of the unit cell are equivalent (Wyckoff site $16f$). The spin-3/2 Co$^{2+}$ atoms are arranged in edge-sharing CoO$_6$ octahedra forming screw chains, running along the $c-$axis, and separated by non-magnetic V$^{5+}$ and Ba$^{2+}$ ions (see Fig.~\ref{StrucNuc}). As schematized by the arrows in Fig.~\ref{StrucNuc}(b), a screw axis $4_1$ runs through the first type of chains of the lattice, transforming Co$_1$ (Co$_{1'}$) at $z=1/8$ into Co$_2$ (Co$_{2'}$) at $z=3/8$, Co$_3$ (Co$_{3'}$) at $z=5/8$, and then Co$_4$ (Co$_{4'}$) at $z=7/8$ (anti-clockwise rotation as $z$ increases), while a screw axis $4_3$ runs through the second type of chain of the lattice, transforming Co$_5$ (Co$_{5'}$) at $z=1/8$ into Co$_6$ (Co$_{6'}$) at $z=3/8$, Co$_7$ (Co$_{7'}$) at $z=5/8$, and then Co$_8$ (Co$_{8'}$) at $z=7/8$ (clockwise rotation as $z$ increases). The Co$_{1'}$...Co$_{4'}$ and Co$_{5'}$...Co$_{8'}$ chains are obtained from the Co$_1$...Co$_4$ and Co$_5$...Co$_8$ chains, respectively, simply by applying the $I-$centering of the lattice: Co$_1$ at $z=1/8$ yields Co$_{3'}$ at $z=5/8$, Co$_5$ at $z=1/8$ yields Co$_{7'}$ at $z=5/8$, ..., by translation of $({1 \over 2},{1 \over 2},{1 \over 2})$.

Two \bacovo single-crystals were grown at Institut N\'{e}el (Grenoble, France) by the floating zone method.\cite{Lejay2011} In both cases, a 5~cm long cylindrical crystal rod, of 3~mm diameter, was obtained, with the growth axis at about 60$^\circ$ from the $c-$axis. From the first rod, a 2.2~mm thick slice was cut perpendicular to the $c-$axis, yielding a 135~mm$^3$ sample with elliptical faces of about $3 \times 6.5$~mm$^2$ (first crystal devoted to the neutron diffraction experiments). Nearby on the same rod, a very small crystal of 2.5 mg was cut with faces perpendicular to the $c-$axis for the specific heat measurements. From the second rod, the sample was cut in order to have a cubic shape, with about 3~mm long edges along the crystallographic axes (second crystal also devoted to the neutron diffraction experiments). Note that this second neutron sample, though 5 times smaller, was much better suited for structure refinements, in particular because of the strong neutron absorption by Co: the various neutron paths (and thus the absorption) were then very similar for all measured reflections. However, the first sample with a much larger volume was more appropriate for the measurements of the incommensurate satellite peaks with small intensity.

\subsection{Specific heat measurements}

The specific heat measurements were performed at CEA-Grenoble\,/\,SPSMS--IMAPEC (Grenoble, France) on a commercial Physical Properties Measurement System (PPMS, Quantum Design) equipped with a $^3$He insert using a semi-adiabatic heat-pulse technique. On this apparatus, magnetic fields up to 9~T could be reached and the temperature was varied between $\sim 600$~mK and 7~K. The 2.5 mg sample was mounted with its cutting surface parallel to the sample platform and thus with the $c-$axis vertical, parallel to the applied magnetic field.

\subsection{Neutron diffraction experimental procedures}

All neutron diffraction experiments were performed at the Institut Laue Langevin (ILL) high-flux reactor in Grenoble, France. The two single-crystals were previously aligned with the $c-$axis vertical on the OrientExpress neutron Laue diffractometer. A series of three experiments were performed on the first single-crystal (with an elliptical shape). The first one was performed on the CSIC/CEA-CRG D15 single-crystal diffractometer, operated in the 4-circle mode, with the sample mounted in a 2~K -- displex refrigerator. Two experiments were then performed on the CEA-CRG D23 single-crystal two-axis diffractometer with a lifting arm detector. The sample was mounted in a 6~T then a 12~T vertical field cryomagnet, equipped with a dilution insert, with the field applied along the $c-$axis. Later on, a fourth experiment was performed on D23 on the second (cubic shaped) single-crystal (not available at the time of the previous experiments). The sample was placed in a standard ILL orange cryostat with the $c-$axis vertical. Incident wavelengths of 1.173~\AA~and 1.280~\AA~were used, respectively on D15 and D23.

For the zero-field experiments on D15 and D23, no collimation was used: only diaphragms were put before and after the sample, in order to optimize the signal over noise ratio. As concerns the experiments performed on D23 with the 6 and 12~T cryomagnets, a 20 arcminutes vertical collimation was put before the sample, in addition to diaphragms, in order to improve the vertical resolution (along the reciprocal lattice vector $\bf c^\star$) and thus to separate the satellites appearing in the LSDW phase at $\pm \xi$ ($\xi << 1$ close to $H_c$). Doing so, a diaphragm of vertical aperture 4 mm could be installed before the detector, without cutting any signal, yielding, together with the use of the collimation, a huge reduction of background. A resolution of about 0.07~r.l.u. (reciprocal lattice units) could then be achieved along the $c^\star$ direction. Putting the $c-$axis of the sample vertical on D23 implied that $hkl$ reflections up to $l=\pm3$ could be collected with the orange cryostat, while ${\bf Q} = (Q_H,\,Q_K,\,Q_L)$ scattering vectors with $-0.5 \leq Q_L \leq 2.15$ and $-0.3 \leq Q_L \leq 1.1$ could be reached with the 6~T and 12~T cryomagnets, respectively.

The zero-field experiment on D15 performed on the first single-crystal was aimed at determining the magnetic structure in the N\'eel phase at $T = 2$~K, together with the nuclear one above and below the transition, and at doing a precise temperature dependence of magnetic peaks. This structure determination was later improved on D23 using the second single-crystal mounted in a cryostat. The experiment done on D23 with the 6~T cryomagnet was aimed at refining the nuclear and magnetic structures in the LSDW phase at $H=4.2$~T and $T = 50$~mK and at measuring temperature and field dependences up to 6~T. These field and temperature scans were then completed at higher field using the 12~T cryomagnet on D23.


\section{Results and analysis}
\label{sec:III}


\subsection{Specific heat measurements}
\label{subsec:IIIA}

\begin{figure}[htb]
	\centering
		\includegraphics[width=8cm]{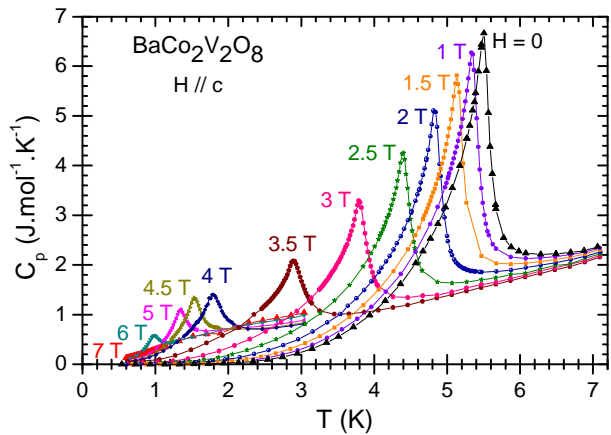}
	\caption{(Color online) Temperature dependence of the specific heat measured in \bacovo in different magnetic fields from 0 to 7~T applied along the $c-$axis.}
	\label{FigCp}
\end{figure}

Before performing any neutron diffraction measurement, our \bacovo single-crystal was first characterized by X-rays\cite{Lejay2011} and specific heat. The aim of the specific heat measurements, performed in a magnetic field applied along the $c-$axis, was to obtain a macroscopic signature of the magnetic-paramagnetic phase boundaries under a magnetic field. The specific heat was measured as a function of the temperature for various magnetic field values ranging between 0 and 9~T (see Fig.~\ref{FigCp}).

In zero magnetic field, a sharp $\lambda-$type anomaly is seen at $T_c \simeq 5.5$~K evidencing a second-order phase transition between the paramagnetic and N\'eel phases when lowering the temperature. The critical temperature $T_c$ and the amplitude of the $\lambda-$like peak both decrease across the paramagnetic-N\'eel phase transition as $H$ increases. A slower decrease of $T_c$ is observed for $H>4$~T, {\it i.e.}, for the paramagnetic-LSDW phase transition. For $H \geq 7$~T, the transition is no longer visible (the measurements performed at 8 and 9~T are not shown), either because the critical temperature is lower than 600~mK (minimal temperature reached during the measurement) or because the anomaly is too weak to be observed. From all the measurements shown in Fig.~\ref{FigCp}, the order-to-disorder transition line in the $H$-$T$ phase diagram was obtained (see Fig.~\ref{DiagHT}) prior to the neutron diffraction measurements.

\subsection{Nuclear structure at $H=0$}
\label{subsec:IIIB}

The first step of the neutron diffraction experiments consisted in refining the low-temperature nuclear structure, previously determined from X-ray diffraction at room temperature.\cite{Wichmann1986} This was done both in the paramagnetic phase and in the N\'eel phase, in order to check that no modification of the nuclear structure occurs when entering the N\'eel phase. To do so, more than 300 independent reflections, allowed in the $I4_1/acd$ space group, were collected both at $T=14$~K and $T=2$~K on D15 using the first crystal. About a hundred independent forbidden reflections were also measured and checked to be null. Though the weighted least-squares $R-$factor was not completely satisfactory ($R_{F^2w} \sim 10\%$), due to the fact that cobalt absorbs neutrons a lot and that this sample has a very asymmetric shape (dimensions of about $6.5 \times 3 \times 2.2$~mm$^3$), these refinements allowed to conclude that the nuclear structure remains identical across the magnetic phase transition, with no change in the atomic coordinates, within the error bars.
      
\begin{table} [htb]
\centering
\caption{Atomic coordinates and isotropic Debye-Waller factors obtained from the nuclear structure refinement performed on D23 on the cubic shaped crystal at $T=1.8$~K in the $I4_1/acd$ space group (agreement $R$ factors: $R_F = 2.38$\% and $R_{F^2w} = 4.75$\%). The larger error bars for the $z$ coordinates come from the fact that the $c-$axis was set vertical on the diffractometer, limiting the $l$ Miller index to $\pm 3$.}
\begin{tabular}{crllll}
\hline
\hline
Atom  & Wyckoff       &     ~$x$        &  $y$        &  $z$         &  $B_{\rm iso}$ (\AA$^2$) \\
      &   site~~~     &                 &             &              &                          \\
\hline
 Ba   &   $8\,a$~~    &  ~   0          &  0.25       &  0.375       &  0.041(21)              \\
 V    &   $16\,e$~~   &  ~   0.08000    &  0          &  0.25        &  0.200                  \\
 Co   &   $16\,f$~~   &  ~   0.16897(5) &  0.41897(5) &  0.125       &  0.398(28)              \\
 O1   &   $32\,g$~~   &  ~   0.15868(3) &  0.07501(3) &  0.38171(14) &  0.301(10)              \\
 O2   &   $32\,g$~~   &  ~   0.49845(2) &  0.08729(3) &  0.34872(12) &  0.247(11)              \\  
\hline
\hline
\end{tabular}
\label{TableRefinement}
\end{table}

We now present in detail the experiment performed on D23 at $T=1.8$~K on the second crystal, which yields a much better refinement owing to its cubic shape. The following lattice parameters were obtained, from centering 22 reflections: $a=12.404(2)$\,\AA~and $c=8.375(12)$\,\AA. 622 nuclear reflections $hkl$ up to $l=\pm 3$ (among them 172 independent ones), allowed in the $I4_1/acd$ space group, were then collected. About 50 independent forbidden reflections $hkl$ with $h+k+l$ even ({\it i.e.}, nuclear reflections allowed by the $I-$centering of the lattice but forbidden either by the screw axis $4_1$ or by a glide plane of the space group) were measured as well and checked to be null. A few $hkl$ reflections with $h+k+l$ odd were also measured at large $Q$ (in order to have a null magnetic contribution) and their zero intensity confirmed the $I-$centering of the lattice. The nuclear refinement was done with the F{\footnotesize ULLPROF} software.\cite{Rodriguez-Carvajal1993} In addition to the scale factor, the $x,y,z$ coordinates and the $B_{\rm iso}$ isotropic Debye-Waller factors were refined for all atoms except vanadium,\cite{vanadium} together with the $\lambda/2$ ratio (smaller than 0.1\%) and the extinction parameters.\cite{extinction} As expected, a much better agreement $R_{F^2w}-$factor than for the first crystal was obtained ($R_{F^2w}=4.75$\%), as a consequence of a better suited crystal shape, yielding the atomic coordinates and isotropic Debye-Waller factors listed in Table~\ref{TableRefinement}. Let us emphasize that the atomic coordinates are very similar to those determined on the first crystal and that the Debye-Waller factors have reasonable values. Last, the four refined extinction parameters are smaller than the six ones refined on the first crystal, which is consistent with the smaller crystal size. The absorption corrections of the intensities [using the A{\footnotesize VEXAR} software from the Cambridge Crystallography Subroutine Library (CCSL)\cite{Brown1981}] did not improve the refinement and was not done in the following magnetic refinements.
 

\subsection{Magnetic structure in the N\'eel phase ($H=0$)}
\label{subsec:IIIC}

The first observation by Kimura {\it et al.}~\cite{Kimura2008b} of the magnetic signal on the 4\,0\,3 reflection allows to easily deduce the propagation vector of the magnetic structure in the N\'eel phase. The magnetic signal appears at $H=0$ at positions $hkl$ with $h+k+l$ odd integer, that is, on the nuclear peaks forbidden by the body-centering $I$ of the lattice. The propagation vector of the magnetic structure is thus ${\bf k}_{AF}=(1,\,0,\,0)$ [or, equivalently, ${\bf k}_{AF}=(0,\,1,\,0)$] owing to the fact that $a>c$.\cite{Rossat-Mignod,rqe2} The magnetic structure was first determined on D15 using the crystal of elliptical shape and then on D23 using the cubic one. Only the latter experiment will be presented again, since both gave the same result with a better agreement factor for the cubic shaped crystal. A total of 313 different magnetic reflections, reducing to 143 independent ones, were collected on D23 on the second crystal, counting 10 seconds per point. The scale factor, the $x$ atomic coordinate and the isotropic Debye-Waller factor of cobalt, the four extinction parameters, and the $\lambda/2$ ratio obtained from the nuclear structure refinement provided us with all the necessary parameters for a quantitative analysis of the magnetic structure. 

For a second-order phase transition (as it is the case here) from the paramagnetic phase to the N\'eel phase, the group theory can be used in order to predict the possible spin structures compatible both with the space group of the nuclear structure and with the propagation vector of the magnetic structure, according to the irreducible representations (irreps) $\tau^\nu$, as described in Ref.~\onlinecite{Bertaut}. As a result, this usually reduces the number of independent parameters to be refined.
The B{\footnotesize ASIREPS} software from the F{\footnotesize ULLPROF} suite was used and four irreducible representations of dimension 2 were found, occurring 3 times each in the group representation: $\Gamma = 3\tau^1 \oplus 3\tau^2 \oplus 3\tau^3 \oplus 3\tau^4$. In the present case, the use of group theory thus reduces the number of fitting parameters from 24 down to 6, whatever the irreducible representation. The four possible magnetic arrangements, deduced from this analysis, are summarized in Table~\ref{GroupTheory}. Note that the Co$_{1'}$...Co$_{8'}$ atoms do not appear in this table. The magnetic moments of the primed cobalt atoms are simply obtained from those of the unprimed ones as follows: the cobalt atoms that are related by the $I$ lattice translation $({1 \over 2},{1 \over 2},{1 \over 2})$ have antiparallel magnetic moments, owing to the propagation vector.

\begin{table} [htb]
\centering
\caption{(Color online) Magnetic components for the eight Co atoms: Co$_1$ (0.331, -0.081, 1/8), Co$_2$ (0.331, 0.081, 3/8), Co$_3$ (0.169, 0.081, 5/8), Co$_4$ (0.169, -0.081, 7/8), Co$_5$ (0.169, 0.419, 1/8), Co$_6$ (0.169, 0.581, 3/8), Co$_7$ (0.331, 0.581, 5/8), and Co$_8$ (0.331, 0.419, 7/8), for each irreducible representation $\tau^1$, $\tau^2$, $\tau^3$, and $\tau^4$. The non-circled signs ('\textcolor{red}{+}' and '\textcolor{red}{$-$}') and the circled signs ('\textcolor{blue}{$\oplus$}' and '\textcolor{blue}{$\ominus$}') correspond to independent magnetic moments.}
\begin{tabular}{cccccccccc}
\hline
\hline
           & ~Co    & ~  1  ~ & ~  2  ~ & ~  3  ~ & ~  4  ~ & ~  5  ~ & ~  6  ~ & ~  7  ~ & ~  8  \\
\hline
           & \textcolor{red}{$m_x$}/\textcolor{blue}{$m'_x$} ~ & ~  \textcolor{red}{+}  ~ & ~  \textcolor{red}{+}  ~ & ~ \textcolor{red}{$-$} ~ & ~ \textcolor{red}{$-$} ~ 
                 & ~ \textcolor{blue}{$\oplus$} ~ & ~ \textcolor{blue}{$\oplus$} ~ & ~\textcolor{blue}{$\ominus$}~ & ~ \textcolor{blue}{$\ominus$} \\
$\tau^1$ ~ & \textcolor{red}{$m_y$}/\textcolor{blue}{$m'_y$} ~ & ~  \textcolor{red}{+}  ~ & ~  \textcolor{red}{$-$}  ~ & ~ \textcolor{red}{$-$} ~ & ~ \textcolor{red}{+} ~ 
                 & ~ \textcolor{blue}{$\oplus$} ~ & ~ \textcolor{blue}{$\ominus$} ~ & ~\textcolor{blue}{$\ominus$}~ & ~ \textcolor{blue}{$\oplus$} \\ 
           & \textcolor{red}{$m_z$}/\textcolor{blue}{$m'_z$} ~ & ~  \textcolor{red}{+}  ~ & ~  \textcolor{red}{$-$}  ~ & ~ \textcolor{red}{+} ~ & ~ \textcolor{red}{$-$} ~ 
                 & ~ \textcolor{blue}{$\oplus$} ~ & ~ \textcolor{blue}{$\ominus$} ~ & ~\textcolor{blue}{$\oplus$}~ & ~ \textcolor{blue}{$\ominus$} \\ 
\hline
           & \textcolor{red}{$m_x$}/\textcolor{blue}{$m'_x$} ~ & ~  \textcolor{red}{+}  ~ & ~  \textcolor{red}{$-$}  ~ & ~ \textcolor{red}{$-$} ~ & ~ \textcolor{red}{+} ~ 
                 & ~ \textcolor{blue}{$\oplus$} ~ & ~ \textcolor{blue}{$\ominus$} ~ & ~\textcolor{blue}{$\ominus$}~ & ~ \textcolor{blue}{$\oplus$} \\ 
$\tau^2$ ~ & \textcolor{red}{$m_y$}/\textcolor{blue}{$m'_y$} ~ & ~  \textcolor{red}{+}  ~ & ~  \textcolor{red}{+}    ~ & ~ \textcolor{red}{$-$} ~ & ~ \textcolor{red}{$-$} ~ 
                 & ~ \textcolor{blue}{$\oplus$} ~ & ~ \textcolor{blue}{$\oplus$}   ~ & ~\textcolor{blue}{$\ominus$}~ & ~ \textcolor{blue}{$\ominus$} \\
           & \textcolor{red}{$m_z$}/\textcolor{blue}{$m'_z$} ~ & ~  \textcolor{red}{+}  ~ & ~  \textcolor{red}{+}    ~ & ~\textcolor{red}{+}    ~ & ~  \textcolor{red}{+}  ~
                 & ~ \textcolor{blue}{$\oplus$} ~ & ~ \textcolor{blue}{$\oplus$}   ~ & ~\textcolor{blue}{$\oplus$}  ~ & ~ \textcolor{blue}{$\oplus$} \\ 
\hline
           & \textcolor{red}{$m_x$}/\textcolor{blue}{$m'_x$} ~ & ~  \textcolor{red}{+}  ~ & ~  \textcolor{blue}{$\oplus$}  ~ & ~\textcolor{red}{+}    ~ & ~  \textcolor{blue}{$\oplus$} ~
                 & ~  \textcolor{red}{$-$}~ & ~  \textcolor{blue}{$\ominus$}~ & ~\textcolor{red}{$-$}  ~ & ~  \textcolor{blue}{$\ominus$} \\
$\tau^3$ ~ & \textcolor{red}{$m_y$}/\textcolor{blue}{$m'_y$} ~ & ~  \textcolor{red}{+}  ~ & ~  \textcolor{blue}{$\oplus$}  ~ & ~\textcolor{red}{+}    ~ & ~  \textcolor{blue}{$\oplus$} ~
                 & ~  \textcolor{red}{+}  ~ & ~  \textcolor{blue}{$\oplus$}  ~ & ~\textcolor{red}{+}    ~ & ~  \textcolor{blue}{$\oplus$} \\
           & \textcolor{red}{$m_z$}/\textcolor{blue}{$m'_z$} ~ & ~  \textcolor{red}{+}  ~ & ~  \textcolor{blue}{$\ominus$}~ & ~\textcolor{red}{$-$}  ~ & ~  \textcolor{blue}{$\oplus$} ~
                 & ~  \textcolor{red}{$-$}~ & ~  \textcolor{blue}{$\oplus$}  ~ & ~\textcolor{red}{+}    ~ & ~  \textcolor{blue}{$\ominus$} \\
\hline
           & \textcolor{red}{$m_x$}/\textcolor{blue}{$m'_x$} ~ & ~  \textcolor{red}{+}  ~ & ~  \textcolor{blue}{$\oplus$}  ~ & ~\textcolor{red}{+}    ~ & ~  \textcolor{blue}{$\oplus$} ~
                 & ~  \textcolor{red}{+}  ~ & ~  \textcolor{blue}{$\oplus$}  ~ & ~\textcolor{red}{+}    ~ & ~  \textcolor{blue}{$\oplus$} \\
$\tau^4$ ~ & \textcolor{red}{$m_y$}/\textcolor{blue}{$m'_y$} ~ & ~  \textcolor{red}{+}  ~ & ~  \textcolor{blue}{$\oplus$}  ~ & ~\textcolor{red}{+}    ~ & ~  \textcolor{blue}{$\oplus$} ~
                 & ~  \textcolor{red}{$-$}~ & ~  \textcolor{blue}{$\ominus$}~ & ~\textcolor{red}{$-$}  ~ & ~  \textcolor{blue}{$\ominus$} \\
           & \textcolor{red}{$m_z$}/\textcolor{blue}{$m'_z$} ~ & ~  \textcolor{red}{+}  ~ & ~  \textcolor{blue}{$\ominus$}~ & ~\textcolor{red}{$-$}  ~ & ~  \textcolor{blue}{$\oplus$} ~
                 & ~  \textcolor{red}{+}  ~ & ~  \textcolor{blue}{$\ominus$}~ & ~\textcolor{red}{$-$}  ~ & ~  \textcolor{blue}{$\oplus$} \\
\hline
\hline
\end{tabular}
\label{GroupTheory}
\end{table}

Looking at Table~\ref{GroupTheory} shows that either the two types of chains (Co$_1$...Co$_4$ and Co$_5$...Co$_8$) or two successive planes along $c$ (at a distance of $\Delta z=1/4$) are magnetically independent, depending on the irreducible representation, $\tau^1$ and $\tau^2$ or $\tau^3$ and $\tau^4$, respectively. Six parameters are thus to be refined in all cases: the magnetic components ($m_x$, $m_y$, $m_z$) of Co$_1$ and ($m'_x$, $m'_y$, $m'_z$) of Co$_5$ or Co$_2$, depending on the irreducible representations we are dealing with, $\tau^1$ and $\tau^2$ or $\tau^3$ and $\tau^4$, respectively. 

The refinement of the magnetic structure, detailed in the following, was tested for each irreducible representation, and only $\tau^1$ gave a satisfactory agreement, with zero $m_x,m'_x,m_y,$ and $m'_y$ (within the error bars). This confirms the prediction of Ising-like chains, with the easy axis along $c$, and that the exchange couplings are antiferromagnetic along the screw chains. The magnetic components along $x$ and $y$ were then fixed to zero.

\begin{figure}[htb]
	\centering
		\includegraphics[width=7cm]{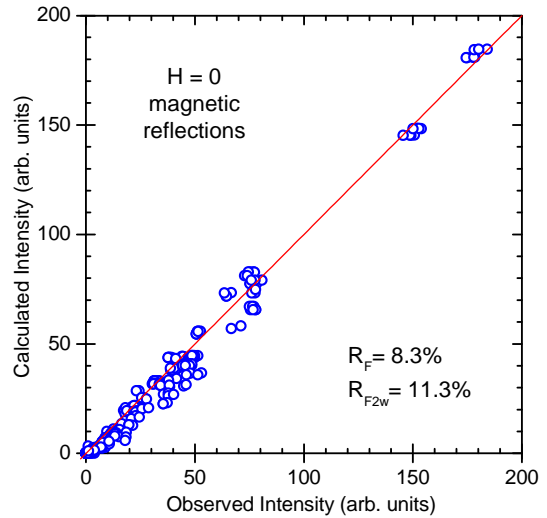}
	\caption{(Color online) Graphical representation of the magnetic structure refinement in the N\'eel phase of \bacovo at $T = 1.8$~K and $H = 0$: calculated versus measured integrated intensities.}
	\label{CalcVsObs}
\end{figure}

Note that Kimura {\it et al.} \cite{Kimura2008b} claimed that the absence of magnetic signal on the $0\,0\,1$ position evidences that the magnetic moments are aligned along the $c-$axis, due to the fact that neutrons are sensitive only to the transverse component (that is, perpendicular to the scattering vector) of the magnetic moment. However, this absence of intensity also occurs for peaks $1\,0\,0$ and $0\,1\,0$, so that no conclusion can be given on the magnetic moment direction from this single observation: the zero intensity in this case rather comes from the magnetic structure factor (magnetic arrangement within the unit cell). A complete refinement was thus necessary, as done in the present work, in order to conclude on the moment orientation.

\begin{table} [htb]
\centering
\caption{Observed intensities $I_{\rm obs}$ of a few magnetic reflections $hkl$, ordered by increasing $\sin \theta/\lambda$, compared to the calculated ones: $I_{\rm calc}^1$ and $I_{\rm calc}^2$ are the contributions of each magnetic domain (with respective populations 49.5\% and 50.5\%) to the total intensity $I_{\rm calc}^{\rm tot}=I_{\rm calc}^1 + I_{\rm calc}^2$. Note that for some of them, only one domain contributes, {\it e.g.}, for $\bar{2}$\,$\bar{3}$\,0 and $\bar{3}$\,$\bar{2}$\,0, while for some others, both domains contribute. The small difference in the total intensity between two reflections $hkl$ and $khl$ is due to the small difference between the domain populations.}
\begin{tabular}{rrrrrrrr}
\hline
\hline
 $h$ & $k$ & $l$ & $\sin \theta/\lambda$ &  $I_{\rm obs}$~~~~ & $I_{\rm calc}^{\rm tot}$~ & $I_{\rm calc}^1$~ & $I_{\rm calc}^2$~ \\
\hline
 1	      & ~~0	       & ~ 0	      & 0.0403   &     0.00(0.23)	&     0.00 &     0.00 &     0.00 \\
 1        & ~~1	       & ~ 1	      & 0.0823   &    24.39(0.37)  &    28.70 &    14.20 &    14.50 \\
$\bar{2}$ & ~~$\bar{3}$ & ~ 0        & 0.1452   &   178.43(1.57)  &   181.15 & ~ 181.15 &     0.00 \\
$\bar{3}$ & ~~$\bar{2}$ & ~ 0        & ~~0.1452 & ~~184.13(1.24)  & ~~184.73 & ~~~~0.00 & ~~184.73 \\
$\bar{2}$ & ~~$\bar{4}$ & ~1	        & 0.1896	 &    32.27(0.64)  &	   33.00 &  	19.87	&    13.13 \\
$\bar{4}$ & ~~$\bar{2}$ & ~1         &	0.1896   &    32.72(0.64)	&    33.15 &  	12.87	&    20.28 \\
 4        & ~~$\bar{3}$ & ~0         & 0.2013   &   145.55(1.25)  &   145.58 &   145.58 &     0.00 \\
 3        & ~~$\bar{4}$ & ~0         & 0.2013   &   150.13(1.26)  &   148.58 &     0.00 &   148.58 \\ 
$\bar{2}$ & ~~$\bar{2}$ & ~ 3        &	0.2116   &     9.55(0.54)  &  	 10.08 &  	 4.99	&     5.09 \\
 5        & ~~$\bar{1}$ & ~$\bar{1}$ & 0.2138	 &	  31.05(0.69)	  &    31.84 &    19.00	&    12.84 \\
 1	      & ~~$\bar{5}$ & ~$\bar{1}$ & 0.2138	 &	  32.82(0.77)	  &    31.97 &    12.57	&    19.40 \\
 5	      & ~~2	       & ~2         & 0.2473   &    19.86(0.83)  &    19.17 &  	19.17	&     0.00 \\
$\bar{2}$ & ~~$\bar{5}$ & ~2	        & 0.2473	 & 	  18.68(1.01)	&    19.56 &     0.00	&    19.56 \\
 3	      & ~~5	       & ~3	        & 0.2949	 & 	  21.93(0.74)	&    19.22 &  	19.22	&     0.00 \\
\hline
\hline
\end{tabular}
\label{IntMag}
\end{table}

The stabilized magnetic structure with the Ising character has a lower symmetry than the paramagnetic space group, and in particular is no longer invariant under the symmetry operator $-y+{1 \over 4},x-{1 \over 4},z+{1 \over 4}$ ($4_1$ screw axis). Two magnetic domains are thus expected, related by this symmetry element, with $m_z=m'_z$ for the first domain and $m_z=-m'_z$ for the second domain, imposing the same value of the magnetic moment on both types of chains. The refinement of the magnetic structure was then performed, for the $\tau^1$ irreducible representation, refining $m_z$ and the domain population. The following results were found, with a weighted least-squares $R-$factor of 11.3\% ($R_F=8.3$\%): $m_z=2.267(3)\mu_B$/Co$^{2+}$ with domain populations of 49.5(4)\% and 50.5(4)\%, that is, equally populated domains, as expected. This moment amplitude of about 2.3$\mu_B$/Co$^{2+}$ is consistent with a magnetic moment of reduced amplitude, as compared to the expected one for spin-3/2 ($m=3\mu_B$), in the presence of quantum effects. The calculated intensities plotted as a function of the observed ones emphasize the quality of the fit (see Fig.~\ref{CalcVsObs}). Note that refining the magnetic structure with a single domain and with no constraint between both types of chains yields, with the same agreement factor and the same calculated intensities, two very different magnetic moments on both chains (3.206 and 0.016$\mu_B$/Co$^{2+}$), which is not realistic.

\begin{figure}[htb]
	\centering
		\includegraphics[width=8cm]{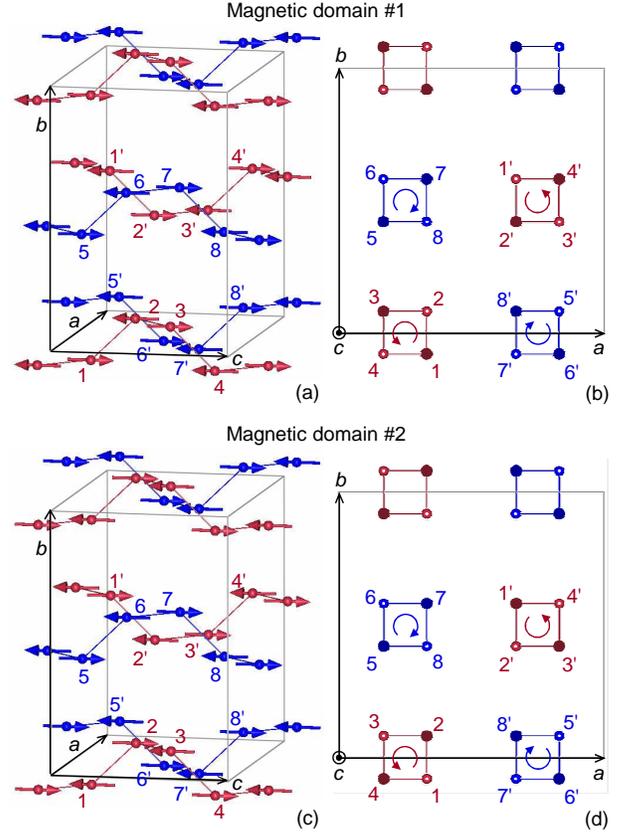}
	\caption{(Color online) Magnetic structure of \bacovo in the two magnetic domains of the N\'eel phase ($H=0$ and $T=1.8$~K): perspective view (left panels) and projection along the $c$-axis (right panels). The cobalt atoms and their magnetic moment are labeled as in Fig.~\ref{StrucNuc}. The two kinds of chains are plotted in different colors: an axis $4_1$ runs through the red chains while an axis $4_3$ runs through the blue ones. Looking at the figures on the right allows an easy visualization of the connection between the two magnetic domains: the application, for example, of the $4_1$ screw axis running along the Co$_1$...Co$_4$ chain transforms domain \#1 into domain \#2 (Co$_1$ transforms into Co$_2$ while Co$_{7'}$ transforms into Co$_{8}$).}
	\label{StrucMagH0}
\end{figure}

The calculated and observed intensities of a few magnetic peaks are listed in Table~\ref{IntMag}. Looking, {\it e.g.}, at reflections $\bar{2}$\,$\bar{3}$\,0 and $\bar{3}$\,$\bar{2}$\,0 shows unambiguously that the presence of both domains is mandatory to explain our data. Figure~\ref{StrucMagH0} shows the magnetic structure in both magnetic domains. In the first domain, the Co$_1$...Co$_4$ chain has magnetic moments parallel to those in the Co$_5$...Co$_8$ chain and antiparallel to those in the Co$_{5'}$...Co$_{8'}$ chain, while it is the contrary in the second domain. One can notice that the $4_1$ screw axis running through the Co$_1$...Co$_4$ chain transforms the Co$_1$--Co$_{7'}$ pair into the Co$_2$--Co$_{8}$ pair [see Fig.~\ref{StrucMagH0}(b)]. Identical interactions couple both pairs which however have a different spin arrangement (antiparallel for Co$_1$--Co$_{7'}$ and parallel for Co$_2$--Co$_{8}$, in the first domain). This suggests the presence of sizable frustration between the two kinds of magnetic chains, which is further argued in Ref.~\onlinecite{Klanjsek}, owing to a detailed analysis of the exchange couplings.

Last, our results compare well with the very recent ones of Kawasaki {\it et al.}\cite{Kawasaki2010} These authors refined the same magnetic structure as the one we obtained for domain \#2, using neutron powder diffraction, which is not sensitive to the domains. Their refined value of $2.18\mu_B$/Co$^{2+}$ for the magnetic moment is slightly smaller than the $2.267(3)\mu_B$/Co$^{2+}$ obtained in the present study.


\subsection{Field and temperature dependences, critical exponents, and $H$-$T$ phase diagram}
\label{subsec:IIID}

\begin{figure}[htb]
	\centering
		\includegraphics[width=8cm]{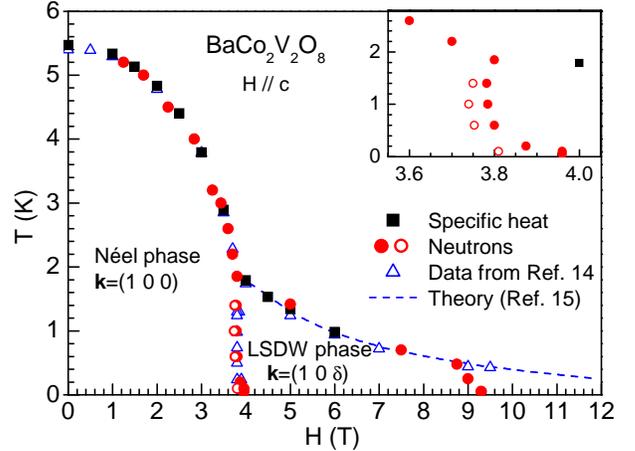}
	\caption{(Color online) Magnetic $H$-$T$ phase diagram of BaCo$_2$V$_2$O$_8$ obtained from the specific heat measurements presented in Sec.~\ref{subsec:IIIA} (black squares) and from the present neutron diffraction measurements (red circles). As concerns the N\'eel-LSDW phase transition (see enlargement in the inset), the solid and open red circles correspond to increasing and decreasing field scans, respectively. The blue open triangles in the main panel are the data of Kimura {\it et al.}~\cite{Kimura2008} obtained from macroscopic measurements and the blue dashed line is the theoretical LSDW-paramagnetic transition line predicted by Okunishi and Suzuki.\cite{Okunishi2007}}
	\label{DiagHT}
\end{figure}

In this section, we present the magnetic field and temperature dependences of selected magnetic reflections, in order to obtain the critical exponents, the field dependence of the propagation vector and of the magnetic intensities, and the phase diagram that is summarized in Fig.~\ref{DiagHT}. Let us recall that when going from the N\'eel to the longitudinal incommensurate magnetic phase (LSDW phase), the antiferromagnetic peak splits above $H_c$ into two incommensurate satellites at positions $\pm \xi$ on either side of the previous peak along $c^\star$ (see the Introduction). Therefore, the propagation vector ${\bf k}_{AF}=(1,\,0,\,0)$ in the N\'eel phase becomes, by continuity, ${\bf k}_{LSDW}=(1,\,0,\,\xi)$, so that the incommensurate satellites are observed at ${\bf H \pm k}_{LSDW}$ with ${\bf H}$ a nuclear scattering vector of the $I$ lattice.\cite{satellites}

\begin{figure}[htb]
	\centering
		\includegraphics[width=8cm]{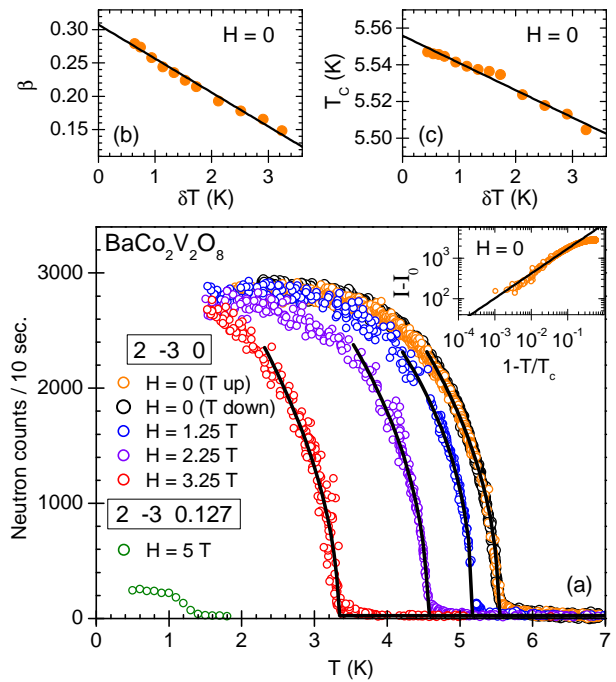}
	\caption{(Color online) (a) Temperature dependence of the magnetic signal: neutron counts on top of the strong 2\,$\bar{3}$\,0 reflection of the N\'eel phase for various magnetic field values between 0 and 3.25~T and on top of the 2\,$\bar{3}$\,0.127 reflection of the LSDW phase at $H=5$ T. The latter curve was obtained from $Q_L-$scans performed across the 2\,$\bar{3}$\,0.127 position counting one minute per point (and then renormalized to 10 seconds for the figure). The black solid lines are power-law fits performed on a temperature range $\delta T=T_c-T \sim 1$~K as explained in the text. (b) and (c) The set of the $\beta$ and $T_c$ values as obtained for various fitting temperature ranges $\delta T$ at $H=0$ together with a linear extrapolation of both parameters to $\delta T=0$ for a precise determination of their values (see text). Inset of panel (a): log-log plot of the magnetic intensity with subtracted background, $I-I_0$, as a function of the reduced temperature, $1-T/T_c$, at $H=0$.  The data (open circles) and the power-law fit (solid line) are presented using the extrapolated values for $\delta T \rightarrow  0$: $T_c = 5.556$~K (for the data and the fit) and $\beta=0.307$ (for the fit).}
	\label{IvsT}
\end{figure}

Figure~\ref{IvsT} (a) presents some temperature scans performed across two different phase transitions: (i) a series of temperature scans performed on top of strong 2\,$\bar{3}$\,0 magnetic peak, across the N\'eel-paramagnetic phase transition, for different field values between 0 and 3.25~T (below $H_c$), (ii) a temperature dependence of the 2\,$\bar{3}$\,$\xi$ reflection with $\xi=0.127$, across the LSDW-paramagnetic phase transition, at $H=5$~T (above $H_c$). The latter curve was obtained from $Q_L-$scans performed across one of the two incommensurate magnetic satellites $\pm \xi$: fitting all this series of scans by a Gaussian function allowed to conclude that the position of the peak does not change at all with temperature and to obtain the temperature dependence of the height of the peak. The transition from the magnetic phases to the paramagnetic one can be clearly seen from the vanishing of the probed reflections. Note that in the LSDW phase, the magnetic signal is much weaker (more than 10 times smaller) than in the N\'eel phase. Last, it is worth noting that the temperature scans performed in zero field for increasing and decreasing temperatures give exactly the same curve, evidencing the absence of any temperature hysteresis for the N\'eel-paramagnetic phase transition and thus in favor of a second-order phase transition.

These measurements allowed to determine the critical exponent $\beta$ for the temperature-induced phase transition from the long-range antiferromagnetic ordering (N\'eel phase) to the paramagnetic phase for different magnetic fields.\cite{extinction2} The order parameter ({\it i.e.}, antiferromagnetic moment $m$) can be expressed as $m(T) \propto (T_c - T)^{\beta}$, so that the neutron intensity curves were fitted to the expression $I(T) = I_0 + A \,(1- T/T_c)^{2\beta}$, where $I_0$ and $A$ are the background intensity and the magnetic intensity at $T=0$, respectively. In our fitting procedure, the background intensity $I_0$ was first determined from the high temperature data and then $A$, $\beta$ and $T_c$ were treated as the fitting parameters. The method, described hereafter, that was used to extract the values of the critical exponent $\beta$ and the critical temperature $T_c$, is the same as the one described {\it e.g.} in Ref.~\onlinecite{Garlea2009}. As exemplified in Figs.~\ref{IvsT}(b) and (c) for the case of $H = 0$, power-law fits to the data were performed over a progressively shrinking temperature range $\delta T = T_c - T$, and the obtained $\beta$ and $T_c$ values were found to increase smoothly, roughly linearly, with decreasing temperature range $\delta T$. The solid lines in Fig.~\ref{IvsT}(a) are such fits for the various temperature scans performed on a temperature range $\delta T \sim 1$~K. Below some critical value $(\delta T)_{\rm min}$ of the fitting temperature range, $\beta$ and $T_c$ were found to suddenly diverge. This divergence is due to the rounding of the transition very close to $T_c$ that prevented us from reducing the fitting range too much. For this reason the set of $\beta$ and $T_c$ values obtained for a set of temperature ranges $\delta T \geq (\delta T)_{\rm min}$ were fitted by a linear regression as a function of $\delta T$, in order to extrapolate the values of $\beta$ and $T_c$ for $\delta T \rightarrow 0$ [see solid line in Figs.~\ref{IvsT}(b) and (c)]. These extrapolated values were considered as the final experimental values of $\beta$ and $T_c$. For the four different fields, no significant variation of the extrapolated $\beta$ value was found (values between 0.307 and 0.328), as expected since we are dealing with the same universality class. These values yield an average critical exponent $\beta = 0.32(1)$, which is consistent with the numerical estimates for the critical exponent of the three-dimensional Ising universality class,\cite{Pelissetto2002} 0.308--0.332, to which \bacovo belongs, and with the one obtained in recent experimental works done in zero field.~\cite{Kawasaki2010,Mansson2012} The critical temperature obtained in zero field by extrapolation [see Fig.~\ref{IvsT}(c)] amounts to $T_c=5.556(2)$~K. To establish the relation of the applied procedure for extracting the values of $\beta$ and $T_c$ to the more familiar one, the zero-field temperature scan together with the corresponding power-law fit are also presented in the following way: log-log plot of the magnetic intensity $I-I_0$ as a function of the reduced temperature $1-T/T_c$ [see inset of Fig.~\ref{IvsT}(a)], where the corresponding extrapolated values of $\beta$ and $T_c$ are used. The power-law fit nicely follows the data up to $1-T/T_c\sim 0.1$ in the log-log representation, which further illustrates the good quality of the fit.

\begin{figure}[h]
	\centering
		\includegraphics[width=8cm]{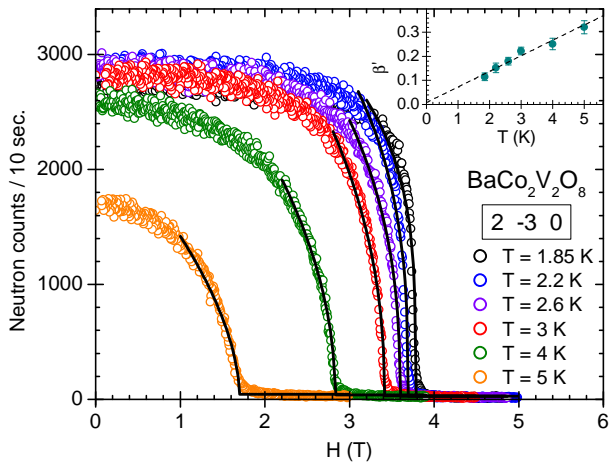}
	\caption{(Color online) Field dependence of the magnetic signal at various temperatures between 1.85~K and 5~K: neutron counts on top of the strong $\bar{2}$\,3\,0 reflection of the N\'eel phase. The black solid lines are fit to a power law as explained in the text. Inset: $\beta'$ values as a function of the temperature (see text). The dashed black line is a linear fit of these data.}
	\label{IvsH}
\end{figure}

Figure~\ref{IvsH} presents some magnetic field scans, performed on top of the 2\,$\bar{3}$\,0 reflection, for different temperatures ranging between 1.85~K and 5~K, across the N\'eel-paramagnetic phase transition. Some of them were done both for increasing and decreasing field (not shown here), resulting  in the absence of any field hysteresis and thus confirming further the prediction of a second-order  N\'eel-paramagnetic phase transition (field-induced order-to-disorder transition of He {\it et al.}\cite{He2005b} mentioned in the introduction). Here again, the same analysis as for the temperature scans was performed, now fitting the neutron intensity curves on a progressively shrinking field range $\delta H = H_c - H$ to: $I(H) = I_0 + A' \,(1- H/H_c)^{2\beta'}$, with $I_0$, $A'$, $H_c$, and $\beta'$ as fitting parameters. The solid lines in Fig.~\ref{IvsH} correspond to such fits performed on a field range $\delta H \sim 0.6$~T. The $\beta'$ values, obtained by extrapolation of the $\beta'(\delta H)$ curves to $\delta H \rightarrow 0$, are plotted in the inset of Fig.~\ref{IvsH} as a function of the temperature. This obtained exponent $\beta'$ decreases in a linear-like fashion from 0.32(3) at $T=5$~K down to 0.12(2) at $T=1.85$~K and extrapolates to $\beta'=0.00(2)$ at $T=0$ (see dashed line in the inset of Fig.~\ref{IvsH}). This result can be understood on the basis of the behavior observed in temperature scans described above, as is discussed in Sec.~\ref{sec:IV}.

\begin{figure}[h]
	\centering
		\includegraphics[width=8cm]{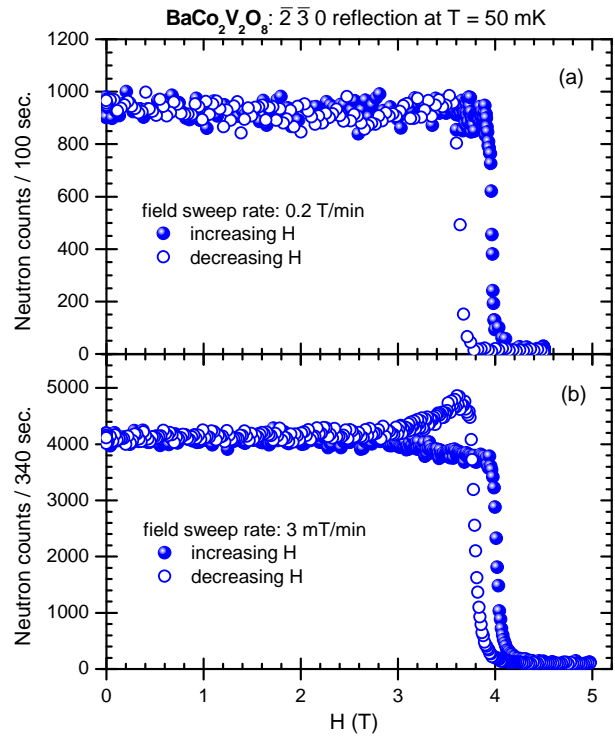}
	\caption{(Color online) Field dependence of the magnetic signal at $T=50$~mK, measured on top of $\bar{2}$\,$\bar{3}$\,0 for two different field sweep rates, 0.2~T/min (a) and 3 mT/min (b), with increasing and decreasing field. A sizable field hysteresis is observed for the N\'eel-LSDW phase transition, whose magnitude increases with the sweep rate ($\Delta H \sim 0.2$~T at 3~mT/min, $\sim 0.32$~T at 0.2~T/min).}
	\label{IvsH3}
\end{figure}

Figure~\ref{IvsH3} presents some magnetic field scans, performed on top of the $\bar{2}$\,$\bar{3}$\,0 magnetic reflection of the N\'eel phase at $T=50$~mK, across the N\'eel-LSDW phase transition, for increasing and decreasing magnetic fields at two different sweep rates. One can notice the presence of a sizable field hysteresis, whose magnitude strongly depends on the field sweep rate, as well as the field dependence of the intensity between about 3~T and 4~T [flat across the complete N\'eel phase for a field sweep rate of 0.2~T/min in Fig.~\ref{IvsH3}(a) but not for a slower field sweep rate in Fig.~\ref{IvsH3}(b)]. This last result is discussed in Sec.~\ref{sec:IV}. All these observations evidence the first-order nature of the N\'eel-LSDW phase transition, consistent with the magnetocaloric measurements of Kimura {\it et al.}\cite{Kimura2008} Such field scans were also performed at different temperatures between 50~mK and 1.8~K (not shown here) in view of completing the phase diagram that is presented in Fig.~\ref{DiagHT}.

\begin{figure}[htb]
	\centering
		\includegraphics[width=8cm]{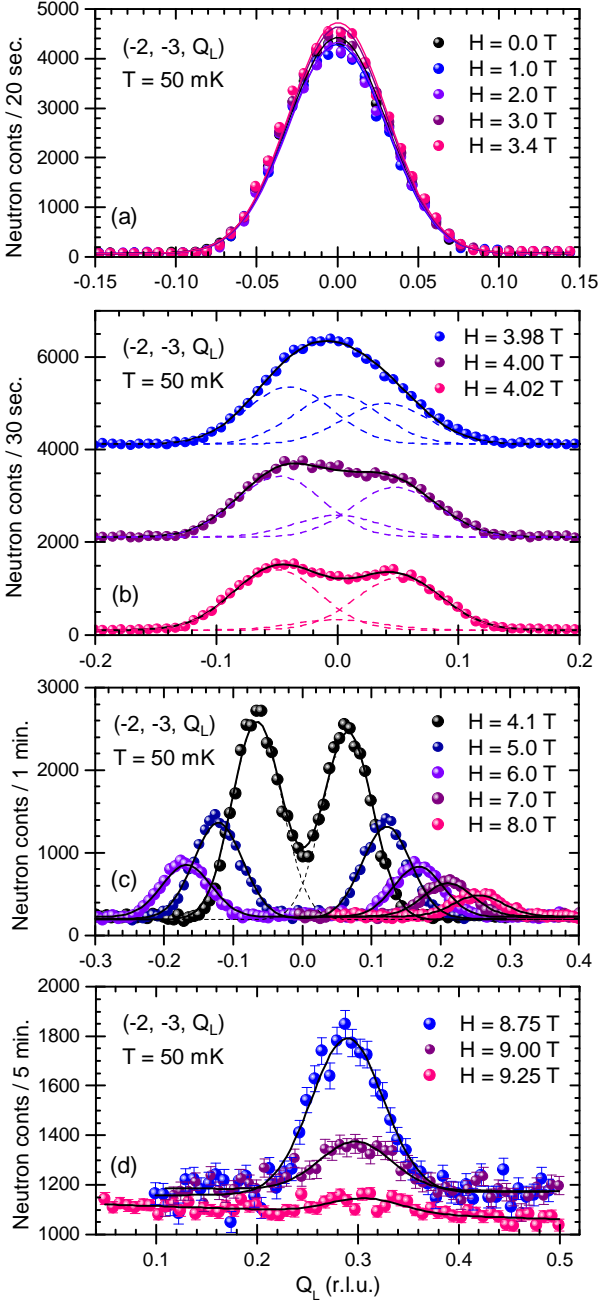}
	\caption{(Color online) $Q_L-$scans performed across the ${\bf Q}=(\bar{2},\bar{3},0)$ scattering vector at $T=50$~mK for different increasing magnetic field values. (a) N\'eel phase, (b) Coexistence of the N\'eel and LSDW phases, (c) and (d) LSDW phase. For a better visibility, the scans measured at 4 and 3.98~T in (b) were shifted vertically by 2000 and 4000, respectively.}
	\label{QL-scans}
\end{figure}

In order to follow the splitting of the antiferromagnetic peak across the N\'eel-LSDW phase transition, $Q_L-$scans (along $c^\star$) were performed centered on the strong  $\bar{2}\,\bar{3}\,0$ magnetic peak at many different field values between 0 and 10~T. Figure~\ref{QL-scans} presents a few of these scans, measured at $T=50$~mK for an increasing magnetic field. For $H \leq 3.92$~T [see Fig.~\ref{QL-scans}(a)], one single magnetic peak is observed in the N\'eel phase: this peak, at every field value, could be well fitted by a Gaussian function located at $Q_L=0$ with half width at half maximum $\delta Q_L = 0.0715$~r.l.u. For $3.94\leq H\leq 4.12$~T [see Fig.~\ref{QL-scans}(b)], two satellites are observed at $Q_L=\pm \xi$ in addition to the previous magnetic peak at $Q_L=0$, evidencing clearly the coexistence of the N\'eel and LSDW phases. This gives an additional strong evidence for the first-order nature of the N\'eel-LSDW phase transition. In this field range, all scans could be well fitted by three Gaussian functions, at $Q_L=0$ and at $Q_L=\pm \xi$. These fits allowed to extract the positions $\pm \xi$ of the satellites of the LSDW phase and the intensity of the three peaks, while their width was fixed to the $\delta Q_L$ value determined from the low field scans. Note that the scans measured between 4.02 and 4.12~T could as well be fitted by two Gaussian functions, {\it i.e.}, with no central peak, but with a larger width. For $H\geq4.14$~T [see Fig.~\ref{QL-scans}(c)], the central peak has completely disappeared, contrary to the observations of Kimura {\it et al.},\cite{Kimura2008b} while the two satellites progressively move away from $Q_L=0$. This discrepancy might come from a different quality of their crystal or a non negligible rate of $\lambda/2$ in their measurements. As can be seen in Figs.~\ref{QL-scans}(c) and (d), a complete scan was performed up to 6~T, fitted by two Gaussian functions symmetric with respect to $Q_L=0$, while only the positive side in $Q_L$ was scanned from 6.5~T (fit by one Gaussian function). The scan at $H=9.25$~T was counted for 10 minutes per point and the satellite peak located at $Q_L=0.308$~r.l.u. is hardly visible. At $H=9.5$~T, nothing was visible after 10 minutes of counting, indicating that the LSDW phase no longer exists above $H_p \simeq 9.25$~T at $T=50$~mK.

\begin{figure}[htb]
	\centering
		\includegraphics[width=8cm]{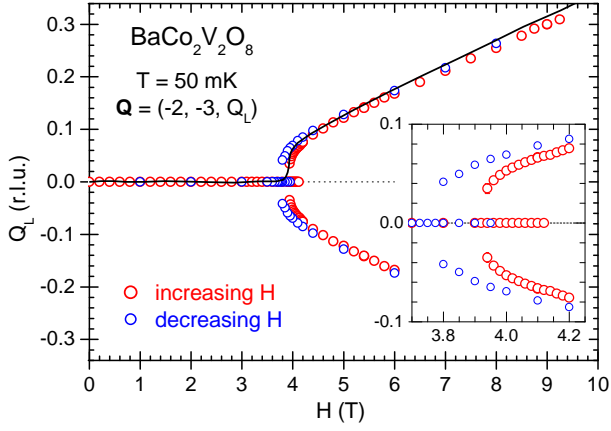}
	\caption{(Color online) Field dependence of the position along $c^\star$ of the magnetic reflections (red and blue circles for increasing and decreasing field, respectively), obtained from the fit of the $Q_L-$scans: antiferromagnetic peak at $Q_L=0$ in the N\'eel phase, incommensurate satellites at $Q_L=\pm \xi$ in the LSDW phase with $\xi$ the incommensurate modulation. The inset presents a enlargement of these curves around the transition. The solid black line in the main panel corresponds to $4 M_z^{\rm cor} / g$, where $M_z^{\rm cor}$ is the magnetization curve $M_z(H)$ of Ref.~\onlinecite{Kimura2008} corrected for a residual Van Vleck contribution~\cite{VV} and $g=6.0$  as described in the text.}
	\label{DeltaVsH}
\end{figure}

\begin{figure}[htb]
	\centering
	\includegraphics[width=8cm]{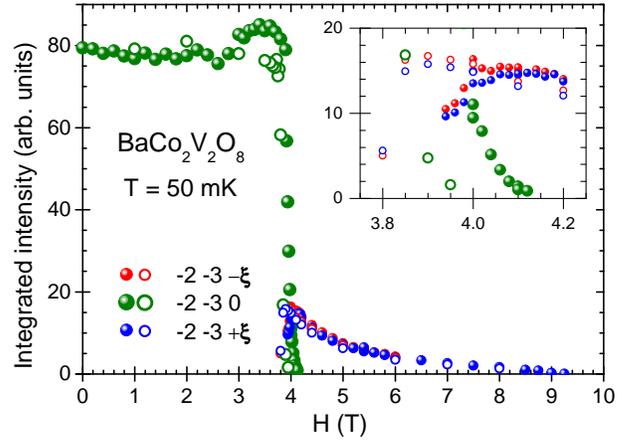}
	\caption{(Color online) Field dependence of the integrated intensity of the magnetic peaks, obtained from the fit of the $Q_L-$scans for increasing (solid circles) and decreasing (open circles) field. The inset presents a enlargement of these data around the transition. The scattering of the red and blue data sets reflects the typical error bars that should be considered.}
	\label{IvsHlowT}
\end{figure}

The previous measurements and analysis were also performed for decreasing field and allowed to obtain precisely the magnetic field dependence of the positions $Q_L$ of the magnetic peaks (see Fig.~\ref{DeltaVsH}). In addition, the field dependence of the integrated intensity was also determined for the magnetic peak of the N\'eel phase, $\bar{2}$\,$\bar{3}$\,0, and for the two satellites of the LSDW phase, $\bar{2}$\,$\bar{3}$\,$\pm \xi$ (see Fig.~\ref{IvsHlowT}).

In Fig.~\ref{DeltaVsH}, a hysteresis of nearly 0.2~T can be seen (emphasized in the inset of Fig.~\ref{DeltaVsH}), as already observed from the field dependence of the signal measured on top of the $\bar{2}\,\bar{3}\,0$ magnetic peak (Fig.~\ref{IvsH3}). This width of nearly 0.2~T corresponds to the field range over which both the N\'eel and the LSDW phases coexist (see inset of Fig.~\ref{DeltaVsH}). The incommensurate modulation of the propagation vector, $\xi$, continuously increases when the field increases, as expected; in other words, it never locks to some commensurate value. Indeed, owing to the fact that there are four Co$^{2+}$ ions along the chain in a unit cell and using Eqs.~(\ref{FermiWaveVector}) and (\ref{PropWaveVector}), it is predicted that the incommensurate modulation of the propagation vector must be proportional to the field-induced uniform magnetization $M_z$ along $c$ as follows: $\xi = 4 \left<S_z\right> = 4 M_z / (g \mu_B)$. In order to compare our results to this prediction, we extracted the magnetization curve from Ref.~\onlinecite{Kimura2008} and further corrected it for a residual Van Vleck paramagnetic contribution of 0.00733 $\mu_B$/T (Ref.~\onlinecite{VV}) to obtain $M_z^{\rm cor}$. Then $4 M_z^{\rm cor} / g = M_z^{\rm cor}/1.5$ was plotted in Fig.~\ref{DeltaVsH} (see black solid line) with $M_z^{\rm cor}$ in $\mu_B$ and using $g=6.0$  as determined from the value of $M_z^{\rm cor}$ at saturation ({\it i.e.}, 3.0$\mu_B$/Co$^{2+}$ at 30 T). The very good agreement between $M_z^{\rm cor}(H)$ and $Q_L(H)=\xi(H)$ in Fig.~\ref{DeltaVsH} confirms the prediction of the theory for the LSDW phase in BaCo$_2$V$_2$O$_8$.

Figure~\ref{IvsHlowT} shows that the integrated intensity of each satellite (LSDW phase) is, just above the transition, about 5 times smaller than the one of the commensurate reflection (N\'eel phase), while it decreases rapidly as the field increases between about 4~T and 9~T, and then completely disappears. Both satellites have exactly the same intensity (within the error bars) on the whole field range as expected, since they are equivalent and have the same $Q$ modulus and thus the same magnetic form factor. Here again, the hysteresis of nearly 0.2~T at the transition is well visible (see the inset of Fig.~\ref{IvsHlowT}).

Last, all the exhaustive temperature and field dependences, like those shown in Figs.~\ref{IvsT}, \ref{IvsH}, \ref{IvsH3}, and \ref{QL-scans}, allowed us to obtain the critical temperatures and critical fields at different magnetic fields and temperatures, respectively, and thus to obtain a detailed $H$-$T$ phase diagram presented in Fig.~\ref{DiagHT} (red circles). For comparison, we add to Fig.~\ref{DiagHT} the points obtained from our specific-heat measurements as well as the points published in  Ref.~\onlinecite{Kimura2008}.

Except for the last two points at 9~T and 9.25~T, our phase diagram is in very good agreement with that of Kimura {\it et al.},\cite{Kimura2008} obtained from macroscopic measurements. The authors measured a higher critical temperature at 9~T than ours and they still observed a transition at 9.5~T, at about 450~mK, while we did not. This discrepancy may arise from a small misalignment of the magnetic field with respect to the chain axis in their experiment, the effect of a magnetic field perpendicular to the chain being much weaker. The enlargement around the N\'eel-LSDW phase transition (shown in the inset of  Fig.~\ref{DiagHT}) also evidences the hysteresis of the transition, as already mentioned, and, surprisingly, a strong curvature of the transition line toward higher field at very low temperatures (below 400~mK). Also, note that the amplitude of the hysteresis is getting larger as the temperature decreases (only 0.05~T around 1~K, about 0.15~T at 50~mK). Last, between $H_c \simeq 3.9$~T and 8.75~T, the agreement between our data and the theoretical prediction given in Ref.~\onlinecite{Okunishi2007} (see dashed blue line in Fig.~\ref{DiagHT}) is excellent, while our experimental results do not follow the theory any more at higher field. Namely, the system leaves the LSDW phase above $H_p\simeq 9.25$~T. The NMR measurements performed by Klanj\v{s}ek {\it et al.}~\cite{Klanjsek} confirm our results with the same critical field $H_p$ and the same transition line for the LSDW phase.


\subsection{Magnetic structure in the LSDW phase ($H=4.2$~T)}
\label{subsec:IIIE}

The magnetic field was set to 4.2~T, coming down from 6~T, at $T=50$~mK, in order to characterize the LSDW phase. These conditions were chosen since they correspond to the best compromise in the LSDW phase between the strongest possible magnetic intensities and the satellites being far-enough apart that they do not overlap at all along $Q_L$. The magnetic peak at $Q_L=0$ (from the N\'eel phase) has also a zero intensity at this magnetic field value. In order to determine precisely the incommensurate modulation $\xi$ of the propagation vector ${\bf k}_{LSDW}=(1,\,0,\,\xi)$ under these conditions, $Q_L-$scans were performed across 12 different pairs of satellites (at $\pm \xi$), well distributed in the reciprocal space. They were then all fitted to extract the position along $c^\star$ of the peak and their average gave $\xi=0.089 (4)$. The magnetic satellites of the LSDW phase thus appear, under these conditions, at ${\bf Q} = {\bf H} \pm {\bf k}_{LSDW} = (h\pm1,~k,~l\pm 0.089)$, with $h$, $k$, $l$ integers satisfying $h+k+l=2n$.

The nuclear structure was first checked to be identical to that in zero field by performing the nuclear structure refinement on a collection of 266 allowed nuclear reflections, reducing to 70 independent ones. This refinement (agreement factor $R_{F^2w}=7.72$\%) also provided us with the necessary information for the magnetic structure refinement ({\it i.e.}, the scale factor, the $x$ coordinate and Debye-Waller factor of Co, the extinction parameters, and the $\lambda/2$ ratio).\cite{absorption} 

\begin{figure}[htb]
	\centering
		\includegraphics[width=7cm]{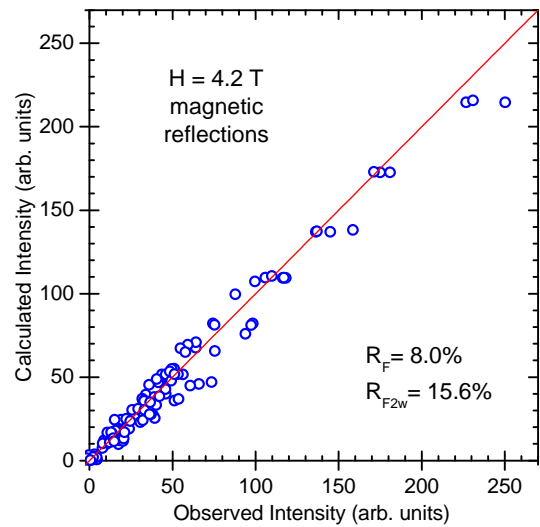}
	\caption{(Color online) Graphical representation of the magnetic structure refinement in the LSDW phase of \bacovo at $T = 1.8$~K and $H = 4.2$~T in decreasing field (coming from 6~T): calculated versus measured integrated intensities.}
	\label{CalcVsObs2}
\end{figure}

\begin{figure*}[htb]
\begin{minipage}[h]{.9\linewidth}
\includegraphics[width=14cm]{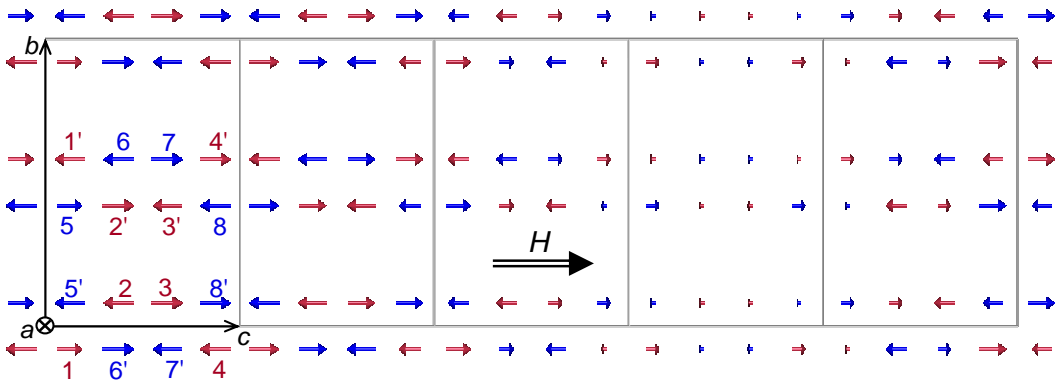}
\end{minipage}
\caption{(color online) Magnetic structure of \bacovo in the LSDW phase at $H=4.2$ T (decreased from 6~T) for the magnetic domain~\#1, shown over about one half period ($\sim 5 c$), using the same labeling and the same colors as in Fig.~\ref{StrucMagH0} for the two types of cobalt chains. This projection on the $(b,c)$ plane is to be compared to Fig.~\ref{StrucMagH0}(a), sketching the magnetic structure of domain \#1 in zero field. The magnetic domain~\#2 of the LSDW phase can simply be obtained from domain \#1 by reverting all spins plotted in red (like for the structure in zero magnetic field).}
\label{StrucMagH4}
\end{figure*}

For the magnetic structure, 129 different magnetic reflections, reducing to 49 independent ones, were collected. The magnetic structure refinement was performed by continuity to what was obtained in zero field, that is: (i) the same structure than in zero field (antiferromagnetic chains with the magnetic moments parallel to the chain $c-$axis and the arrangement between chains as shown in Fig.~\ref{StrucMagH0}) but with an additional modulation (sine wave function) of the moment amplitude along the chains, and (ii) the presence of two domains, corresponding to each other by the $4_1$ screw axis, as described for the N\'eel magnetic structure. In a first refinement, two parameters were fitted, the amplitude $A$ of the sine wave function and the domain population, and the following was found: $A=1.395(6)\mu_B$ with populations 38.6(8)\% and 61.4(8)\%, with the following agreement $R-$factors: $R_F=7.97$\% and $R_{F^2w}=15.6$\%. \cite{absorption,intensity} Note that the maximum magnetic moment in the LSDW phase is smaller than the $\sim 2.3 \mu_B$ of the N\'eel phase. The plot of the calculated intensities as a function of the observed ones (see Fig.~\ref{CalcVsObs2}) emphasizes the quality of the refinement.

Figure~\ref{StrucMagH4} shows the magnetic structure for domain \#1 (using the same labeling as for the zero-field magnetic structure), projected on the $(b, c)$ plane over about one half period along the chain axis ($\sim 5 c$). The non-equirepartition of the domain populations arising from the refinement seems surprising and is discussed in Sec.~\ref{sec:IV}. A second refinement was then performed, imposing equal domain populations (50\% each) and the following was found: a very similar moment amplitude $A=1.364(8)\mu_B$ with (as expected) worse $R-$factors: $R_F=13.4$\% and $R_{F^2w}=25.8$\%. As a result, the correct analysis of the domain population is not crucial for the magnetic moment amplitude, since very similar moment amplitude values are obtained in the two refinements. In conclusion, this magnetic structure refinement confirms for the first time that: (i) the magnetic moment amplitudes of the four cobalt atoms along the chain (at $z=$1/8, 3/8, 5/8, and 7/8) follow the global sine wave function imposed by the incommensurate modulation $\xi=0.089$ of the propagation vector, thus, the magnetic ordering is indeed a spin density wave, and (ii) the magnetic moments are lying along the $c-$direction, hence this spin density wave is longitudinal.

\begin{figure}[htb]
	\centering
		\includegraphics[width=8cm]{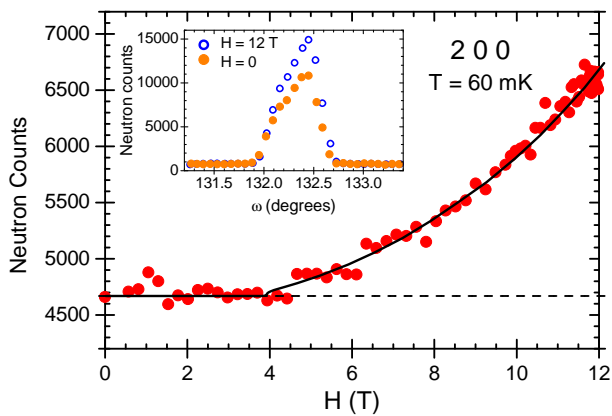}
	\caption{(Color online) Count on top of the 2\,0\,0 reflection as a function of the magnetic field evidencing the ferromagnetic magnetization $M_z$ arising when the field increases. The dashed horizontal black line corresponds to the intensity $I_0$ on top of the pure nuclear reflection. The solid black curve corresponds to the square of the magnetization curve $M_z$ of Ref.~\onlinecite{Kimura2008} corrected for a residual Van Vleck contribution,~\cite{VV} scaled to our data, and with an offset $I_0$. Inset: $\omega-$scan on the 2\,0\,0 reflection at $H = 0$ (pure nuclear reflection) and $H=12$~T (nuclear and ferromagnetic reflection) allowing a precise determination of the ferromagnetic moment at 12~T.}
	\label{Mferro}
\end{figure}

In addition to the longitudinal incommensurate component discussed above, a ferromagnetic component is induced by the applied magnetic field (field-induced uniform magnetization $M_z$), corresponding to the one obtained in macroscopic magnetization measurements. The latter can also be determined from neutron diffraction, as described in the following. The field dependence of the square of the uniform magnetization was obtained on the complete accessible magnetic field range (0~\nobreakdash--~12~T) by counting on top of the 2\,0\,0 weak nuclear reflection, for which the ferromagnetic component is strong (see Fig.~\ref{Mferro}). Qualitatively, these measurements agree perfectly well with the magnetization measurements of Kimura {\it et al.},\cite{Kimura2008} which are also represented in the figure after subtraction of the residual Van Vleck contribution~\cite{VV} and after the appropriate treatment for a scaling to our data. For a quantitative comparison, rocking curves with a long counting time were then performed at 60~mK on the 2\,0\,0 reflection at $H=0$ and $H=12$~T (see the inset of Fig.~\ref{Mferro}). Their difference corresponds to the ferromagnetic contribution estimated to 0.68$\mu_B$/Co$^{2+}$ using the scale factor obtained from structural refinement. This result is consistent with the magnetization measurements of Kimura {\it et al.}:\cite{Kimura2008} at $T = 1.3$~K and $H=12$~T, the authors measured $M_z \sim 0.78\mu_B$/Co$^{2+}$yielding $M_z^{\rm cor} \sim 0.69\mu_B$/Co$^{2+}$ after correction for the residual Van Vleck contribution.~\cite{VV}

           
\section{Discussion}
\label{sec:IV}

The experimental data described in Sec.~\ref{sec:III} reveal some features that deserve to be further discussed. Let us first discuss the consequences of the existence of magnetic domains, which was probed in this single-crystal neutron diffraction study. The magnetic domains at $H=0$ were found to be equally populated, as expected. However, a sizable non-equirepartition of the domains was found in the LSDW phase (populations $39\%\,/\,61\%$) at 4.2~T, as mentioned in Sec.~\ref{subsec:IIIE}. This could be due to the shape of the crystal (very elongated along one direction) combined with the strong absorption of cobalt. However, then the refinement of the zero-field magnetic structure, also performed on this crystal, would have yielded the same domain populations, which is not the case.

The non-equirepartition of the two magnetic domains in this field-induced LSDW phase at $H=4.2$~T is rather due to a small misalignment of the applied magnetic field of the 6~T vertical cryomagnet with respect to the $c-$axis, yielding an asymmetrization of both domains.\cite{UBmatrix} Namely, the $a-$axis was tilted by 1.5$^\circ$ from the horizontal plane while the $b-$axis was tilted only by 0.6$^\circ$. As a consequence, the applied vertical field was at 88.5$^\circ$ from the $a-$axis and at 89.4$^\circ$ from the $b-$axis, instead of 90$^\circ$ from both of them, yielding a nearly three times larger field component along the $a-$axis than along the $b-$axis (0.11~T / 0.04~T, for a 4.2~T applied vertical field), and thus favoring one domain with respect to the other. Such a large sensitivity of the domain populations to the field misalignment was also observed in the magnetoelectric compound MnPS$_3$.\cite{Ressouche2010}

This non-equirepartition of the magnetic domains upon a sizable magnetic field could also explain the behaviors of the field scans in Fig.~\ref{IvsH3} (performed in the same experiment) and Fig.~\ref{IvsHlowT} (performed with the 12~T vertical cryomagnet), as developed hereafter. The field dependence of the intensity of the magnetic $\bar 2$\,$\bar 3$\,$0$ reflection suggests that the populations of both magnetic domains do not vary at all when the sweep rate is high [see Fig.~\ref{IvsH3}(a)], while they do vary in the $\sim 1$~T wide range below the transition for a low sweep rate [see Fig.~\ref{IvsH3}(b)]. Indeed, let us recall that this reflection is sensitive only to the first magnetic domain (see Table~\ref{IntMag}). The only other possible explanation for the field scans of Fig.~\ref{IvsH3}(b) would be a sudden variation of the magnetic moment in the 1~T range below the transition, which seems unrealistic and which was never predicted. The small bump observed in this field range for decreasing $H$ is also present in Fig.~\ref{IvsHlowT}, which was obtained on the same magnetic reflection at an even slower sweep rate. Indeed, the tilt of the sample, in this experiment using the 12~T cryomagnet, was very similar (1.2$^\circ$ for the $a-$axis, 0.5$^\circ$ for the $b-$axis).

Next, we discuss the experimentally obtained critical exponents of the order parameter $m$ ({\it i.e.}, the antiferromagnetic moment) for the transition from the N\'eel phase to the paramagnetic phase above $1.85$~K. As this is a second-order phase transition, the critical exponents characterizing the transitions in the temperature direction and in the field direction are expected to be the same away from the multicritical points.~\cite{Paul1990} For our case, this can easily be shown if we write
\begin{equation}
m(H,T) = m_0 \left[ 1-\frac{T}{T_c(H)} \right]^\beta,
\label{critical}
\end{equation}
where $T_c(H)$ is the shape of the phase boundary, $\beta$ is the critical exponent in the temperature direction and $m_0$ is the zero-temperature order parameter. We assume that both $\beta$ and $m_0$ are field-independent, which is compatible with the experimental results presented in Fig.~\ref{IvsT}. Taking into account that the $T_c(H)$ dependence can be locally, for small variations of $H$, approximated by a linear function, it is easy to show analytically that the critical exponent $\beta'$ determined by varying $H$ at constant $T$ must be the same, $\beta' = \beta$. In contrast to this, we experimentally found $\beta'$ to amount to the value of $\beta$ only in the upper part of the boundary, at $5$~K, whereas we found it to decrease in a roughly linear fashion with decreasing temperature, extrapolating to zero at zero temperature, as summarized in the inset of Fig.~\ref{IvsH}. Such a behavior may be an artifact of the method we used to experimentally extract the critical exponent $\beta'$. Namely, the method that combines fitting to a power law in a progressively shrinking field range with the extrapolation to the phase boundary, leads to the correct value of the critical exponent only when $\beta'$ depends on the width of the selected field range in a  linear way. If the exponent $\beta'$ reaches the value of $\beta$ only very close to the phase boundary, in the range where experimental fits are no longer feasible, the method apparently yields a wrong value. As shown in the Appendix, this is exactly the case, so that the observed temperature dependence of the extracted $\beta'$ is a mere experimental artifact.

Now, we wish to discuss the $H$-$T$ phase diagram we obtained and to compare it with the theoretical predictions, in the light of the recent NMR results from Klanj\v{s}ek {\it et al.}~\cite{Klanjsek} First, our refined magnetic structure at 4.2~T fully confirms for the first time the realization in \bacovo of the predicted LSDW phase, in contrast to the previous neutron diffraction study of Kimura {\it et al.},\cite{Kimura2008b} which demonstrated only that the magnetic order is incommensurate. Second, the first-order character of the N\'eel-LSDW phase transition, first proposed by Kimura {\it et al.}\cite{Kimura2008} from field hysteresis in their magnetocaloric measurements, was confirmed by our neutron diffraction measurements: (i) a sizable field hysteresis of the transition, (ii) a sweeping field rate dependence of both the magnetic intensities and the amplitude of the field hysteresis, (iii) the coexistence of the N\'eel and the LSDW orderings over a sizable field range. The first-order character of the N\'eel-LSDW phase transition has not been discussed theoretically to our knowledge.

Last, the LSDW-paramagnetic transition line in our $H$-$T$ phase diagram perfectly coincides, up to 8.75~T, with that obtained experimentally by Kimura {\it et al.}\cite{Kimura2008} and theoretically by Okunishi and Suzuki.~\cite{Okunishi2007} Some deviation from these works is however observed at higher fields, in agreement with the NMR measurements of Klanj\v{s}ek {\it et al.}~\cite{Klanjsek} From our neutron measurements and this NMR study, the critical temperature of the LSDW phase suddenly drops down to zero when the magnetic field is increased from 8.75 to $H_p \simeq 9.25$~T. This drop corresponds to the onset of a new type of magnetic ordering,\cite{Klanjsek} not predicted by Okunishi and Suzuki,\cite{Okunishi2007} that occurs before the expected transverse staggered ordering~\cite{Okunishi2007} at higher fields ($H > H^* \sim 15~$T). The NMR observations in this new phase were explained involving a new type of order (named ferromagnetic SDW in their work), where the chains with LSDW are ferromagnetically coupled both along the $a-$axis and along the $b-$axis, with a long-wavelength amplitude modulation in the $(a,b)$ plane.~\cite{Klanjsek} Describing the interchain exchange couplings, the authors were able to develop a model that reproduces both the LSDW (named columnar SDW in their work) and ferromagnetic SDW orders. Following these NMR observations and their interpretation, the ferromagnetic SDW phase was searched in our last neutron diffraction experiment at $H=12$~T and $T=50$~mK, by scanning the reciprocal lattice around positions compatible with this phase. Unfortunately, no signal was observed and additional measurements are thus needed.




\section{Conclusion}
\label{sec:V}
The field-induced magnetic properties of the Ising-like spin chain system \bacovo have been studied in detail by means of single-crystal neutron diffraction. Our study was performed with a careful analysis of the magnetic domains distribution, allowing a full determination of the magnetic arrangements in both the N\'eel and LSDW phases, at $H=0$ and 4.2~T, respectively. The determination of the magnetic structure in the incommensurate phase at 4.2 T is the first direct evidence of this predicted LSDW ordering in \bcv. The field-induced transitions from the N\'eel to the paramagnetic phase and to the LSDW phase at lower temperature were studied by following the positions and intensities of chosen magnetic reflections up to 12~T and down to 50~mK. The phase boundary was found to be first-order for the N\'eel-LSDW transition and second-order for the N\'eel-paramagnetic transition, where critical exponents could be extracted from the field and temperature scans, confirming that \bacovo belongs to the 3D Ising universality class. A complete investigation of the $H$-$T$ phase diagram as well as the field-dependence of the magnetic propagation vector was performed and found to compare perfectly to the TLL predictions up to 8.75~T. A departure from the prediction is observed above 8.75~T where the LSDW magnetic satellites disappear. This is well below the expected transition to the transverse staggered long-range order, a discrepancy already pointed out by NMR studies and calling for further investigations.


\section{Acknowledgments}
We thank L.-P. Regnault, E. Ressouche, and M. Enderle for fruitful discussions, F. Bourdarot and D. Braithwaite for their help in the specific heat measurements, P. Fouilloux and B. Vettard for their technical support during the neutron diffraction measurements, and J. Rodr\'iguez-Carvajal for his help in the magnetic structure refinement. Part of this work has been supported by the French ANR project NEMSICOM.


\section{APPENDIX}

\begin{figure}[htb]
	\centering
		\includegraphics[width=8cm]{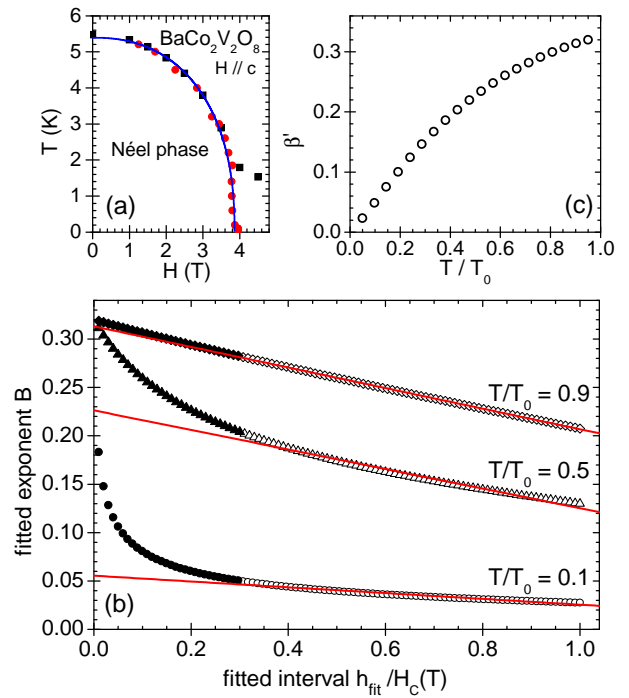}
	\caption{(Color online) (a) Enlargement on the N\'eel phase of the $H$-$T$ phase diagram. The same symbols as in Fig.~\ref{DiagHT} are used for the experimental data. Phase boundary is very well approximated by the shape of the superellipse (blue line) as described in the text, (b) Dependence of the ``critical'' power-law exponent obtained in a fitting procedure as a function of the width of the fitting interval, simulated for field sweeps at various constant temperatures. As the values close to zero are experimentally inaccessible, a linear extrapolation (red lines) of the available points (open symbols) to zero clearly underestimates the critical value, with progressively stronger underestimate as the temperature is lowered. (c) Temperature dependence of the obtained ``critical'' exponent $\beta'$, to be compared with the inset of Fig.~\ref{IvsH}.}
	\label{BetaPrime}
\end{figure}

Assuming that Eq.~(\ref{critical}) is valid for temperature sweeps across the phase boundary, with $m_0$ and $\beta$ field-independent, we show that the critical exponent $\beta'$  experimentally extracted from the field sweeps across the boundary is apparently temperature-dependent. In order to do that, we convert the magnetization given by Eq.~(\ref{critical}) into a form appropriate for the field sweeps. For simplicity we take the phase boundary with the shape of the superellipse, $T_c(H) = T_0 \sqrt[n]{1-(T/T_0)^n}$, which approximates the experimental boundary very well for $n=2.32$, $T_0=5.39$~K and $H_0=3.86$~T [see Fig.~\ref{BetaPrime}(a)]. In addition, we switch to the variable measuring the departure from the boundary in the field direction, $h = H_c(T)-H = H_0 \sqrt[n]{1-(T/T_0)^n} - H$, where $H_c(T)$ defines the transition field at a given temperature $T$. This leads to
\begin{equation}
m(h,T) = m_0 \left[ 1-\frac{T/T_0}{\sqrt[n]{1 - \left[ \sqrt[n]{1-(T/T_0)^n} - h/H_0 \right]^n}} \right]^{\beta}.
\label{critical2}
\end{equation}

For the relevant $\beta = 0.32$ value and a given constant value of $T$, this $h$-dependence of $m$ is numerically fit, by a least squares fit, to a ``critical'' power law, $A(h/H_c)^B$, in the chosen interval $0 < h < h_{\textrm{fit}}$. We plot thus obtained critical exponent $B$ as a function of the fitting range in dimensionless units, $h_{\textrm{fit}}/H_c(T)$, as shown in Fig.~\ref{BetaPrime}(b). This procedure is identical to the one used in the experimental determination of the critical exponent $\beta'$, where $\beta'$ is found by a linear extrapolation of $B[h_{\textrm{fit}}/T_c(H)]$ to zero. The main difference is that in this simulation of the experimental procedure, due to the absence of any noise and distortions of the data, we have a proper access to the required $h_{\textrm{fit}} \rightarrow 0$ limit. Indeed, in Fig.~\ref{BetaPrime}(b) we see that mathematically correct $B(0) = \beta = 0.32$ limit is reached at any temperature, but in a progressively narrower fit range $h_{\textrm{fit}}/H_c(T)$ as the temperature is lowered.

In contrast to that, very narrow $h_{\textrm{fit}}/H_c(T)$ ranges are experimentally inaccessible because the fit becomes unstable. Taking into account that in a typical experimental fitting we follow $B[h_{\textrm{fit}}/H_c(T)]$ dependence in a range from $h_{\textrm{fit}}/H_c(T)=1$ down to only $0.3$, corresponding to open symbols in Fig.~\ref{BetaPrime}(b), we see that a linear extrapolation of these points to zero, $B_{\textrm{exp}}(0)$, gives the correct value $B_{\textrm{exp}}(0) = \beta$ only at high temperature. As the temperature is lowered, $B_{\textrm{exp}}(0)$ is suppressed in a monotonous manner, tending to zero at zero temperature [see Fig.~\ref{BetaPrime}(c)], reproducing closely the experimentally observed behavior displayed in the inset of Fig.~\ref{IvsH}.


\bibliographystyle{apsrev}

\begin{references}
\bibitem{vanadium} Note that for the vanadium atom, the $x$ coordinate (not fixed on the $16e$ Wyckoff site) and the isotropic Debye-Waller factor were not refined, since vanadium is purely incoherent for neutron.
\bibitem{extinction} The phenomenological model for single-crystal refinements was used, involving six extinction parameters along the crystallographic axes and their bisectors. However, owing to the tetragonal symmetry of the space group and to the crystal shape (same dimension along $a$ and $b$), the following constraints were used for the extinction parameters: $Ext_a=Ext_b$ and $Ext_{a+c}=Ext_{b+c}$.
\bibitem{extinction2} The extinction was checked to have a negligible variation across the transition, so that it does not affect the temperature and field scans of the magnetic reflections presented in Figs.~\ref{IvsT} and \ref{IvsH}.
\bibitem{absorption} Note that the agreement is not as good as for the cubic crystal in zero field (see Secs.~\ref{subsec:IIIB} for the nuclear structure, \ref{subsec:IIIC} for the magnetic structure), because of the asymmetric shape of this crystal (elliptical instead of cubic) which affects the equivalent reflections, due mostly to the strong absorption of cobalt.
\bibitem{intensity} An additional reason for the poorer $R_F-$factor of the LSDW phase refinement as compared to the N\'eel phase are the 7 times smaller intensities of the magnetic reflections in the former.
\bibitem{UBmatrix} Using the UB matrix obtained from the centering of about 30 reflections in the experiment using the 6~T vertical cryomagnet (in which the LSDW magnetic structure was refined and the field scans of Fig.~\ref{IvsH3} were measured), the precise orientation of the $a$ and $b$ axes with respect to the vertical magnetic field could be calculated.
\end{references}

\end{document}